\documentclass[12pt,epsf,amssymb]{article}

\usepackage{graphicx}
\usepackage{axodraw}
\usepackage{amsfonts}

\begin{document}
\baselineskip 0.6cm
\newcommand{\gsim}{ \mathop{}_{\textstyle \sim}^{\textstyle >} }
\newcommand{\lsim}{ \mathop{}_{\textstyle \sim}^{\textstyle <} }
\newcommand{\vev}[1]{ \left\langle {#1} \right\rangle }
\newcommand{\bra}[1]{ \langle {#1} | }
\newcommand{\ket}[1]{ | {#1} \rangle }
\newcommand{\Dsl}{\mbox{\ooalign{\hfil/\hfil\crcr$D$}}}
\newcommand{\nequiv}{\mbox{\ooalign{\hfil/\hfil\crcr$\equiv$}}}
\newcommand{\nsupset}{\mbox{\ooalign{\hfil/\hfil\crcr$\supset$}}}
\newcommand{\nni}{\mbox{\ooalign{\hfil/\hfil\crcr$\ni$}}}
\newcommand{\EV}{ {\rm eV} }
\newcommand{\KEV}{ {\rm keV} }
\newcommand{\MEV}{ {\rm MeV} }
\newcommand{\GEV}{ {\rm GeV} }
\newcommand{\TEV}{ {\rm TeV} }

\def\diag{\mathop{\rm diag}\nolimits}
\def\tr{\mathop{\rm tr}}
\def\diag{\mathop{\rm dim}\nolimits}

\def\Z{\mathbb{Z}}
\def\R{\mathbb{R}}
\def\C{\mathbb{C}}
\def\Im{\mathrm{Im}}
\def\Re{\mathrm{Re}}
\def\vol{\mathrm{vol}}

\def\Spin{\mathop{\rm Spin}}
\def\SO{\mathop{\rm SO}}
\def\O{\mathop{\rm O}}
\def\SU{\mathop{\rm SU}}
\def\U{\mathop{\rm U}}
\def\Sp{\mathop{\rm Sp}}
\def\SL{\mathop{\rm SL}}


\begin{titlepage}

\begin{flushright}
RESCEU-3/04 \\
LBNL-54540 \\
UT04-03 \\
\end{flushright}

\vskip 2cm
\begin{center}
{\large \bf Product-Group Unification in Type IIB String Theory}

\vskip 1.2cm
Taizan Watari${}^a$ and T.~Yanagida${}^b$

\vskip 0.4cm
${}^a$ 
{\it Department of Physics, University of California at Berkeley,\\ 
          Berkeley, CA 94720, USA} \\
${}^b$
{\it Department of Physics, University of Tokyo, \\
          Tokyo 113-0033, Japan}\\

\vskip 1.5cm
\abstract{The product-group unification is a model of 
unified theories, in which masslessness of the two Higgs doublets 
and absence of dimension-five proton decay are guaranteed 
by a symmetry.  
It is based on SU(5) $\times$ U($N$) ($N=2,3$) gauge group. 
It is known that various features of the model are explained 
naturally, when it is embedded in a brane world. 
This article describes an idea of 
how to accommodate all the particles of the model in Type IIB 
brane world. The GUT-breaking sector is realized by a D3--D7 system, 
and chiral quarks and leptons arise from intersection of D7-branes. 
The D-brane configuration can be a geometric realization 
of the non-parallel family structure of quarks and leptons, 
an idea proposed to explain the large mixing angles observed 
in the neutrino oscillation.
The tri-linear interaction of the next-to-minimal supersymmetric 
standard model is obtained naturally in some cases. 
}

\end{center}
\end{titlepage}


\section{\label{sec:Intro}Introduction}

Supersymmetric grand unified theories (SUSY GUTs) have been 
considered as the most attractive framework beyond the standard model, 
ever since the gauge-coupling unification was observed \cite{SU(5)}.
However, the GUT scale ($\sim 10^{16}$ GeV) is much higher than 
the energy frontier, and only a little has been known about 
how the theory should be.

The two following clues are the most important among all.
First, the two Higgs doublets of the minimal SUSY standard model
(MSSM) should be almost massless, although their mass term 
is not forbidden by the gauge group of the standard model. 
Second, dimension-five operators contributing to proton decay 
\cite{dim5-op} should be sufficiently suppressed 
\cite{dim5-exp,dim5-clcB}, although they are not 
forbidden either. 
These two clues are not independent. 
Suppose that either of those operators is forbidden by a symmetry, 
and then the other is also forbidden;
\begin{equation}
W \; \nni \; HH \qquad \longleftrightarrow \qquad 
W \; \nni \; \psi \psi \psi \psi,
\end{equation} 
provided the Yukawa coupling of quarks and leptons ($\psi$) with 
the Higgs particles ($H$) 
\begin{equation}
W \ni \psi \psi H
\end{equation}
is allowed by the symmetry.
Thus, the two clues can be considered as manifestation of 
an unbroken symmetry.

In the presence of such a symmetry, only three types of 
mass matrices are possible for the Higgs particles.
One is the missing VEV (vacuum expectation value) type \cite{DW}.
Models with the unbroken symmetry were obtained in 
\cite{BDS,Barr,WittenG2,DNS}.
Another is the missing partner type \cite{mpm}.
Models were obtained in \cite{Yanagida,HY,IY}.
The last type involves infinite number of Higgs particles, 
which can be obtained as a Kaluza--Klein tower 
\cite{CHSW,WittenE8,HN}.

We investigate the model \cite{Yanagida,HY,IY} 
of the missing partner type. 
It was pointed out in \cite{IWY} that various features 
of the model are naturally explained when it is embedded 
in a brane world. 
Thus, Ref.~\cite{WY01,WY02} tried to embed the D = 4 model 
into toroidal orientifolds of the Type IIB theory, although 
the variety of toroidal orientifolds was not enough 
to accommodate all the matter contents of the model, i.e., 
the particles in the GUT-symmetry-breaking sector, Higgs multiplets, 
and quarks and leptons. 
One of the purposes of this article is 
to illustrate an idea of how to embed all the particles 
in the Type IIB string theory.
We do not restrict ourselves to toroidal orientifolds, and consider 
generic (orientifolded)  Calabi--Yau 3-folds.
Various properties of the D = 4 model are translated into 
local properties of D-brane configuration and geometry 
of the Calabi--Yau 3-folds \cite{IIB-local}. 
We also extract some implications to phenomenology, 
although an explicit string model is not obtained 
in this article.

This article is organized as follows. 
We briefly review the D = 4 model \cite{Yanagida,HY,IY}
in section \ref{sec:Review}, 
and explain the motivations to embed it in a brane world. 
The model is embedded in the Type IIB theory in section \ref{sec:IIB}.
Subsection \ref{ssec:GUT} describes the embedding of the sector 
responsible for breaking the unified symmetry. 
Subsection \ref{ssec:Higgs} is devoted to Higgs particles, where 
we also see that 
the next-to-minimal SUSY standard model (NMSSM) 
\cite{NMSSM} is obtained as a special case.
Subsection \ref{ssec:QL} describes the idea of how to obtain 
quarks and leptons.
The origin of the $B-L$ symmetry and right-handed neutrinos are 
also discussed.
The D-brane configuration there provides geometric 
realization of the non-parallel family structure \cite{SY,Ramond,HMW}, 
which is suggested by the large mixing angles in the neutrino 
oscillation. 
The final section is devoted to summary and open questions.

Note: Errors in the expressions of Ramond--Ramond charge are 
corrected\footnote{We thank Kentaro Hori for useful comments.} 
and two paragraphs are newly added (one in section 3.3.2 and 
the other in the appendix B) in version 2, in March, 2006.

\section{\label{sec:Review}Brief Review of Previous Results}

\subsection{\label{ssec:model}Review of the Model}
Product-group unification models based on product gauge groups, 
SU(5)$_{\rm GUT}\times$U(3)$_{\rm H}$ and 
SU(5)$_{\rm GUT}\times$U(2)$_{\rm H}$, are constructed
in a four-dimensional (D = 4) spacetime with ${\cal N}$ = 1 SUSY 
\cite{Yanagida,HY,IY}. They are quite similar.
Thus, we only review the SU(5)$_{\rm GUT}\times$U(3)$_{\rm H}$ model 
in this section. 
For a review of the SU(5)$_{\rm GUT}\times$U(2)$_{\rm H}$ model, 
see \cite{WY02,IW}.

The ``unified gauge symmetry'' (or which we refer 
to as the GUT symmetry in the following) of the model 
is SU(5)$_{\rm GUT}\times$U(3)$_{\rm H}$ instead of the simple SU(5).
Quarks and leptons are singlets of the U(3)$_{\rm H}$ gauge group 
and form three families of {\bf 5}$^*$+{\bf 10} representations 
of SU(5)$_{\rm GUT}$.
The Higgs multiplets that contain the two Higgs doublets are 
$H({\bf 5})^i$ and $\bar{H}({\bf 5}^*)_i$, 
which are also singlets of U(3)$_{\rm H}$.
Fields introduced to break the GUT symmetry are the followings:
$X^{\alpha}_{\;\;\beta}(\alpha ,\beta =1,2,3)$ transforming 
as (${\bf 1}$,{\bf adj.}=${\bf 8}+{\bf 1}$) under 
the SU(5)$_{\rm GUT}\times$U(3)$_{\rm H}$ gauge group, 
and $Q^{\alpha}_{\;i}(i=1,\ldots,5)+Q^{\alpha}_{\;6}$
and $\bar{Q}^i_{\;\alpha}(i=1,\ldots,5)+\bar{Q}^6_{\;\alpha}$
transforming as ({\bf 5}$^*$+{\bf 1},{\bf 3}) and 
({\bf 5}+{\bf 1},{\bf 3}$^*$).
The SU(5)$_{\rm GUT}$ indices are denoted by $i,j$ and those of 
U(3)$_{\rm H}$ by $\alpha,\beta$.
The chiral superfield $X^\alpha_{\;\;\beta}$ is also written as 
$X^c (t_c)^\alpha_{\;\;\beta}(c=0,1,\ldots,8)$, 
where $t_a(a=1,\ldots,8)$ are Gell-Mann matrices 
of the SU(3)$_{\rm H}$ gauge group\footnote{
The normalization condition $\tr(t_a t_b) = \delta_{ab}/2$ 
is understood. 
Note that the normalization of the following $t_0$ is determined 
so that it also satisfies $\tr(t_0t_0)=1/2$.} and 
$t_0 \equiv {\bf 1}_{3 \times 3}/\sqrt{6}$,
where U(3)$_{\rm H}\simeq$SU(3)$_{\rm H}\times$U(1)$_{\rm H}$.

The superpotential of the model is given \cite{IY} by 
\begin{eqnarray}
W & = & 
   \sqrt{2} \lambda_{\rm 3H} 
      \bar{Q}^i_{\;\alpha} X^a(t_a)^{\alpha}_{\;\beta}Q^{\beta}_{\;i} 
 + \sqrt{2} \lambda_{\rm 3H}' 
      \bar{Q}^6_{\;\alpha} X^a(t_a)^{\alpha}_{\;\beta}Q^{\beta}_{\;6}  
                                  \nonumber \\
  &   &       \!\!\!\!\! 
 + \sqrt{2} \lambda_{\rm 1H} 
      \bar{Q}^i_{\;\alpha} X^0(t_0)^{\alpha}_{\;\beta}Q^{\beta}_{\;i} 
 + \sqrt{2} \lambda_{\rm 1H}' 
      \bar{Q}^6_{\;\alpha} X^0(t_0)^{\alpha}_{\;\beta}Q^{\beta}_{\;6}
                                  \nonumber \\
  &   &       \!\!\!\!\!
 - \sqrt{2}\lambda_{\rm 1H}  v^2 X^\alpha_{\;\alpha}  
                                  \label{eq:super}  \\
  &   &       \!\!\!\!\! 
 + h' \bar{H}_i \bar{Q}^i_{\;\alpha} Q^{\alpha}_{\;6} 
 + h  \bar{Q}^6_{\;\alpha} Q^{\alpha}_{\;i} H^i   
                                  \nonumber \\
  &   &       \!\!\!\!\! 
 + y_{\bf 10} {\bf 10} \cdot {\bf 10} \cdot H 
 + y_{\bf 5^*} {\bf 5}^* \cdot {\bf 10} \cdot \bar{H} + \cdots,
                                   \nonumber
\end{eqnarray}
where the parameter $v$ is taken to be of the order of the GUT scale, 
$y_{\bf 10}$ and $y_{\bf 5^*}$ are Yukawa coupling constants 
for the quarks and leptons, and $\lambda_{\rm 3H},\lambda_{\rm 3H}',
\lambda_{\rm 1H},\lambda_{\rm 1H}',h'$ and $h$ 
are dimensionless coupling constants. 
Ellipses stand for mass terms of the neutrinos and 
(other) non-renormalizable terms. 
The fields $Q^{\alpha}_{\;i}$ and $\bar{Q}^i_{\;\alpha}$ 
acquire vacuum expectation values (VEV's), 
$\vev{Q^\alpha_{\;\;i}}$ = $v \delta^\alpha_{\;\; i}$ and 
$\langle\bar{Q}^i_{\;\;\alpha}\rangle$ = $v \delta^i_{\;\;\alpha}$, 
because of the second and third lines in (\ref{eq:super}). 
Thus, the gauge group $\SU(5)_{\rm GUT}\times \U(3)_{\rm H}$ 
is broken down to that of the standard model.
Mass terms of the coloured Higgs multiplets arise from 
the fourth line in (\ref{eq:super}) in the GUT-breaking vacuum. 
On the other hand, mass terms of the Higgs doublets are absent 
in the superpotential (\ref{eq:super}), and hence they are massless.
No unwanted particle remains in the massless spectrum 
after the GUT symmetry is broken down 
to SU(3)$_C\times$SU(2)$_L\times$U(1)$_Y$.

The superpotential (\ref{eq:super}) has a (mod 4)-$R$ symmetry;  
the $R$ charges (mod 4) of the fields are given 
in Table \ref{tab:u3-R4}.
\begin{table}
 \caption{\label{tab:u3-R4} $R$ charge assignment of 
the SU(5)$_{\rm GUT}\times$U(3)$_{\rm H}$ model}
\begin{center}
 \begin{tabular}{c|ccccccc}
  Fields & ${\bf 5}^*_i$, ${\bf 10}^{ij}$ & $H^i$ & $\bar{H}_i$ 
         & $X^\alpha_{\;\beta}$ 
         & $Q^{\alpha}_{\;i}$, $\bar{Q}^i_{\;\alpha}$ 
         & $Q^\alpha_{\; 6}$ & $\bar{Q}^6_{\; \alpha}$ \\
\hline
  $R$ charge (mod 4) & $1$ & $0$ & $0$ & $2$ & $0$ & $2$ & $-2$ \
 \end{tabular}
\end{center}
\end{table}
This symmetry forbids various dangerous operators such as 
mass terms of the Higgs doublets $W \supset \bar{H}_i H^i$ and 
proton-decay operators of dimension-four 
$W \supset {\bf 5}^* \cdot {\bf 10} \cdot {\bf 5}^*$ 
and dimension-five 
$W \supset {\bf 5}^* \cdot {\bf 10}\cdot {\bf 10}\cdot {\bf 10}$.
It is not broken even after the GUT symmetry is broken, 
because the VEV $\vev{\bar{Q}Q}$ does not carry the $R$ charge.
It is broken down to the $R$ parity when the SUSY is broken. 
Therefore, the two Higgs doublets have a mass term 
(so called $\mu$-term) only after the SUSY is broken.
The dimension-five proton-decay operators 
have a suppression factor $(m_{\rm SUSY}/M_*)$, 
and hence they are irrelevant to the proton decay. 
Thus, the two difficulties in SUSY GUTs discussed 
in the introduction are naturally solved by the (mod 4)-$R$ symmetry

It is also remarkable \cite{KMY} that the discrete $R$ symmetry 
is free from a  mixed anomaly ($R$ mod 4)[SU(3)$_{\rm H}$]$^2$, and 
can be free from another mixed anomaly 
($R$ mod 4)[SU(5)$_{\rm GUT}$]$^2$.
Therefore, the symmetry can be an unbroken subgroup of 
a gauge symmetry. 
This fact sheds a light on a question why the (mod 4)-$R$ symmetry 
is preserved in an accuracy better than $10^{-14}$. 

The fine structure constants of the MSSM are given 
at the tree level by
\begin{equation}
\frac{1}{\alpha_{3}} \equiv 
 \frac{1}{\alpha_{C}} = \frac{1}{\alpha_{\rm GUT}} 
                                       + \frac{1}{\alpha_{\rm 3H}}, 
\label{eq:treematchU3-3}
\end{equation}
\begin{equation}
\frac{1}{\alpha_{2}} \equiv
 \frac{1}{\alpha_{L}} =  \frac{1}{\alpha_{\rm GUT}}, 
\quad \quad \quad 
\label{eq:treematchU3-2}
\end{equation}
and 
\begin{equation}
\frac{1}{\alpha_{1}} \equiv
\frac{\frac{3}{5}}{\alpha_{Y}} = \frac{1}{\alpha_{\rm GUT}} + 
                                  \frac{\frac{2}{5}}{\alpha_{\rm 1H}},
\label{eq:treematchU3-1}
\end{equation} 
where $\alpha_{\rm GUT}$, $\alpha_{3\rm H}$ and $\alpha_{1\rm H}$ 
are the fine structure constants of SU(5)$_{\rm GUT}$, 
SU(3)$_{\rm H}$ and U(1)$_{\rm H}$, respectively.
Thus, the MSSM coupling constants
$\alpha_{3}$, $\alpha_{2}$ and $\alpha_{1}$ are unified approximately, 
when $\alpha_{3\rm H}$ and $\alpha_{1\rm H}$ are sufficiently large.
Although the unification is no longer a generic prediction 
of the present model, 
it is a consequence of the relatively large gauge coupling
constants $\alpha_{\rm 3H}$ and $\alpha_{\rm 1H}$, or equivalently, 
of the relatively small constant $\alpha_{\rm GUT}$.
The condition for the approximate unification is 
\begin{equation}
 \frac{1}{\alpha_{\rm GUT}} \gsim 
(10\mbox{--}100)\times (\frac{1}{\alpha_{\rm 3H}},~\frac{1}{\alpha_{\rm 1H}}).
\label{eq:disparity}
\end{equation}

Now we have two remarks on the present model. First, 
let us neglect the weak coupling SU(5)$_{\rm GUT}$ interactions, 
and then the first three lines of the superpotential (\ref{eq:super}) 
preserves ${\cal N}$ = 2 SUSY. 
Indeed, the chiral multiplet $X^\alpha_{\;\; \beta}$ 
is identified with the ${\cal N}$ = 2 SUSY partner 
of the U(3)$_{\rm H}$ vector multiplet, and the chiral multiplets 
$Q+\bar{Q}$ are regarded as ${\cal N}$ = 2 hypermultiplets. 
The first two lines in the superpotential are nothing 
but a part of gauge interactions 
of the ${\cal N}$ = 2 SUSY theory when 
\begin{equation}
 \alpha_{\rm 3H} \simeq \alpha_{\rm 3H}^{\lambda^{(')}} \equiv
\frac{\lambda_{\rm 3H}^{(')2}}{4\pi}, {\rm ~~and~~}
 \alpha_{\rm 1H} \simeq \alpha_{\rm 1H}^{\lambda^{(')}} \equiv
\frac{\lambda_{\rm 1H}^{(')2}}{4\pi}, 
\label{eq:N=2relation}
\end{equation}
are satisfied. 
The third line is interpreted as the Fayet--Iliopoulos F-term.
Now, this partial ${\cal N}$ = 2 SUSY is not only theoretically 
interesting, 
but also plays important roles in phenomenology: 
(i) the renormalization-group flow of $\alpha_{\rm 3H}$ 
is stabilized only when the ${\cal N}$ = 2 SUSY relation 
(\ref{eq:N=2relation}) is preserved approximately, 
and (ii) large threshold corrections from the particles 
in the SU(3)$_C$-{\bf adj.} representation 
vanish when it is preserved.
For more details, see \cite{FW,IW}.

The second remark is that the cut-off scale $M_*$ should be lower 
than the Planck scale $M_{\rm pl} \simeq 2.4 \times 10^{18}$ GeV. 
There are two reasons for this.
First, the coupling constant $\alpha_{\rm 1H}$ becomes too large 
below the Planck scale because the U(1)$_{\rm H}$ interaction 
is asymptotically non-free (see \cite{FW,IW} for more details).
The second reason is that $\vev{\bar{Q}Q}/M_*^2$, 
which breaks the SU(5)$_{\rm GUT}$ symmetry, 
should not be too small, or otherwise, it would be unable 
to account for the difference between the Yukawa coupling constants 
of the strange quark and of the muon at the unification scale.

\subsection{\label{ssec:mot}
Motivations to Embed the Model into String Theory}

References \cite{IWY,WY01,WY02} tried to embed the D = 4 models  
in subsection \ref{ssec:model} into a brane world.
This subsection briefly explains the motivations of the embedding. 
One may skip this subsection, because the construction 
in string theory begins in the next section.

The product-group unification model in the previous subsection 
preserves ${\cal N}$ = 2 SUSY in the GUT-breaking sector.
However, the full theory has only ${\cal N}$ = 1 SUSY. 
Thus, it is a logical possibility that the multiplets 
${\cal W}$ and $X^\alpha_{\;\;\beta}$ are not 
related by anything like ${\cal N}$ = 2 SUSY in their origin, 
neither are $Q$ and $\bar{Q}$,  
and that their interactions look like those of ${\cal N}$ = 2 
gauge theories accidentally.
However, there are two important observations against this 
possibility. 
First, the GUT-breaking sector 
(with the ${\cal N}$ = 2 SUSY) couples only 
to the SU(5)$_{\rm GUT}$ gauge fields and 
to the Higgs multiplets. 
There is no direct coupling between the sector and the chiral quarks 
and leptons.
Thus, the GUT-breaking sector is decoupled from other sectors 
(with only ${\cal N}$ = 1 SUSY) when only a few relatively weak 
couplings are turned off, and then, ``symmetry of the sector'' 
is well-defined.
Second, the ${\cal N}$ = 2 SUSY in the GUT-breaking sector 
plays important roles in phenomenology \cite{FW,IW}.
Therefore, the apparent ${\cal N}$ = 2 SUSY in the sector 
can be a symmetry of more fundamental theory, rather than 
an accidental symmetry.
 
It is not easy to understand the coexistence of the ${\cal N}$ = 2 
and ${\cal N}$ = 1 SUSY in D = 4 theories, 
but easy in theories based on higher-dimensional spacetime.
Higher-dimensional theories have extended SUSY. 
They are compactified on curved manifolds so that D = 4 theories 
with only ${\cal N}$ = 1 SUSY are obtained at low energies.
Let us assume that there is a point in the internal manifold 
around which 
an extended SUSY such as ${\cal N}$ = 2 is preserved, 
while the full geometry has only ${\cal N}$ = 1 SUSY. 
If the GUT-breaking sector is localized at such an ${\cal N}$ = 2 
preserving area, then, the particle contents and interactions between 
them satisfy the ${\cal N}$ = 2 SUSY, as in the model.
This is the primary motivation to embed the product-group unification 
model into a brane-world;  
the GUT-breaking sector are supposed to be localized on ``branes''.

In string theories, $N$ coincident D-branes support U($N$) 
gauge theories on their world volume. 
Thus, the product gauge group is quite ubiquitous in string vacua 
with D-branes. 
In particular, five coincident D-branes and $N$
coincident D-branes realize U(5)$_{\rm GUT}\times$U($N$)$_{\rm H}$ 
gauge group\footnote{The 
extra U(1) gauge symmetry contained 
in U(5)$_{\rm GUT}\times$U(3)$_{\rm H}$ can be identified 
with the $B-L$ symmetry. 
It requires right-handed neutrinos so that 
its triangle anomaly is cancelled. 
Right-handed neutrinos can lead to tiny masses of left-handed 
neutrinos through the see-saw mechanism 
\cite{see-saw}. See discussion in subsubsection \ref{ssec:QL}.} 
($N=2,3$). 
Massless fields in the bi-fundamental representations appear 
on the common locus of the two stacks of D-branes.
References \cite{IWY,WY01,WY02} consider that 
the U($N$)$_{\rm H}$-charged particles, i.e., the GUT-breaking sector,
is obtained on a world volume of D3-branes, and that 
the SU(5)$_{\rm GUT}$ gauge field comes from D7-branes.
Rotational symmetry of the internal space 
plays the role of the $R$ symmetry.

When the total volume of the internal space  is large,  
the fundamental scale $M_*$ 
is lower than the Planck scale $M_{\rm pl}$. 
This is one of the features required for  
the SU(5)$_{\rm GUT}\times$U($N$)$_{\rm H}$ model ($N=2,3$).
Moreover, the world volume of the D-branes for the U(5)$_{\rm GUT}$ 
gauge group can also be large since the total volume is large.
When the world volume is moderately large, 
i.e., 10--100 in units of $1/M_*$,  
the fine structure constants of the U(5)$_{\rm GUT}$ interactions 
are relatively small by 1/10--1/100 when compared with those of 
U(3)$_{\rm H}$. 
This is also a desired feature of the model.

In summary, the string vacua with D-branes may be able to explain 
five features\footnote{It was also pointed out in \cite{IWY} that 
the SUSY flavour problem may be solved through the gaugino-mediation 
mechanism \cite{gaugino} in this framework. 
This is because the SU(5)$_{\rm GUT}$ gauge field propagates 
in the internal dimensions, say, on the D7-branes, while 
quarks and leptons may be further localized 
inside the world volume of SU(5)$_{\rm GUT}$  
(see subsection \ref{ssec:QL}).
After \cite{IWY} was published, however, 
phenomenological and theoretical problems of the gaugino mediation 
were pointed out by \cite{gaugino-old-phen} and 
\cite{gaugino-old-theor}, respectively. 
Further investigation is necessary on this issue 
\cite{gaugino-new-phen,gaugino-new-theor}.} 
of the SU(5)$_{\rm GUT}\times$U($N$)$_{\rm H}$ model 
($N=2,3$), namely, 
(i) the partial ${\cal N}$ = 2 SUSY in the GUT-breaking sector, 
(ii) the origin of the product gauge group, 
(iii) the origin of the $R$ symmetry, 
(iv) the cut-off scale lower than the Planck scale
and (v) the relatively small coupling constant 
of the SU(5)$_{\rm GUT}$ interactions. 
Therefore, we consider that it is well motivated to embed 
the model into string vacua.

\section{\label{sec:IIB}Construction in Type IIB Theory}

We consider that the SU(5)$_{\rm GUT} \times$U($N$)$_{\rm H}$ 
gauge symmetry ($N=2,3$) arises from space-filling D-branes. 
The Type IIB theory is compactified on Calabi--Yau 3-folds, 
and space-filling D-branes and orientifold planes 
wrap holomorphic cycles \cite{BBS,OOY}, so that  
D = 4 gauge theories with ${\cal N}$ = 1 SUSY 
are obtained at low energies.
Quarks and leptons arise from open strings connecting those D-branes.
This is the picture we have in mind.

Reference \cite{IIB-local} emphasized that the discovery of 
D-branes brought a new method into string phenomenology.
Provided that a gauge theory is obtained from D3-branes 
trapped at an orbifold singularity, some of 
(though not necessarily all of) phenomenological properties 
of the gauge theory are determined only from the local 
geometry and local configuration of the D-branes.
One does not have to know the whole information of the Calabi--Yau 
3-fold in obtaining (some of) phenomenological predictions. 
References \cite{G2-localA,WittenG2,G2-localB} and \cite{IIA-local} 
share the same philosophy in a slightly extended form; 
geometry around singularities and some properties of cycles 
on which Kaluza--Klein monopoles and D6-branes are wrapped 
are sometimes sufficient information  
in deriving some of phenomenological consequences.
This approach can be called as local construction or 
bottom-up approach.

This approach enables us to proceed just as particle physicists 
have been doing.
Each sector of a phenomenological model is realized by 
D-branes. Only minimal requirement is imposed on the local geometry 
so that the desired properties of the model are obtained. 
After that, the local configuration of D-branes for various sectors 
are combined in a suitable way to form the whole model.
One can leave any properties in such a construction, if 
one does not have to fix them. 
Yet, one can hope to obtain some phenomenological consequences.
This is the approach we adopt in this article.  

\subsection{\label{ssec:GUT}GUT-Breaking Sector}

\begin{figure}[tb]
 \begin{center}
\begin{picture}(200,100)(0,0)
   \resizebox{5cm}{!}{\includegraphics{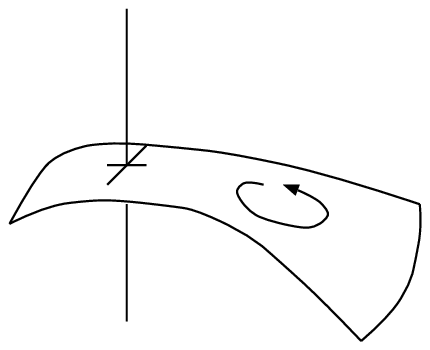}} 

   \Text(-101,61)[c]{$\bullet$}
   \Text(-70,90)[cb]{locally {\bf C}$^2$/{\bf Z}$_M$ $\times$ {\bf C}}
   \LongArrow(-70,88)(-95,65)

   \Text(10,30)[l]{D7's wrapped on a 4-cycle $\Sigma_1$}
   \LongArrow(15,25)(-5,15)
   \Text(-130,75)[rb]{fract. D3's}
   \LongArrow(-145,72)(-105,63)
   \Text(-55,20)[rt]{non-trivial Chan--Paton bundle}
   \LongArrow(-75,25)(-55,42)
\end{picture}
 \caption{A schematic picture of the local geometry of $CY_3$ 
and D-brane configuration in it.
Only the parts relevant to subsection \ref{ssec:GUT} (GUT-breaking
  sector) and \ref{ssec:Higgs} (Higgs particles) are described here.}
\label{fig:3+6}
 \end{center}
\end{figure}

$N$ fractional D3-branes ($N=2,3$) and six D7-branes are 
introduced for the GUT-breaking sector. 
The local geometry around the fractional D3-branes 
(the GUT-breaking sector) is ${\bf C}^2/{\bf Z}_M \times {\bf C}$.
The fractional D3-branes are at the orbifold singularity, 
and the D7-branes are stretched in ${\bf C}^2/{\bf Z}_M$ 
direction while not in ${\bf C}$ direction (see Fig.~\ref{fig:3+6}).
Then, D = 4 ${\cal N}$ = 2 SUSY is preserved in the field theory 
localized at the fractional D3-branes, 
due to the local D-brane configuration and geometry.

When the $N$ fractional D3-branes belong to the same representation 
of the orbifold group ${\bf Z}_M$, they are trapped at 
the orbifold singularity. 
We have a U($N$) vector multiplet of ${\cal N}$ = 2 SUSY 
from the D3--D3 open strings, while massless hypermultiplets 
from the D3--D3 strings are projected out.
If the six D7-branes are in the same representation of ${\bf Z}_M$ 
as the $N$ fractional D3-branes are, 
then the massless hypermultiplets from the D3--D7 open strings 
are not projected out from the spectrum.
In summary, we have an ${\cal N}$ = 2 SUSY U($N$) 
gauge theory of D = 4 spacetime, 
whose matter contents are six hypermultiplets 
in the fundamental representation.
The flavour symmetry of this hypermultiplets is U(6), which 
arises from the D7-branes.
The matter content is just what we want for the GUT-breaking sector.
Thus, we identify the U($N$) vector multiplet and hypermultiplets 
with the U($N$)$_{\rm H}$ vector multiplet 
(${\cal W}$,$X^\alpha_{\;\;\beta}$) and the hypermultiplets 
($Q^\alpha_{\; k}$,$\bar{Q}^k_{\;\;\alpha}$) 
in the bi-fundamental representation ({\bf $N$},{\bf 5}$^*$+{\bf 1}) 
(or equivalently, in  ({\bf $N$}$^*$,{\bf 5}+{\bf 1})) 
in the previous section. 

The centre-of-mass U(1) part of the U($N$)$_{\rm H}$ 
vector multiplets, however, is no longer massless 
at the quantum level \cite{fract-mass}. 
Thus, there should be at least one more fractional D3-brane 
at the {\bf C}$^2$/{\bf Z}$_M$ singularity. 
This extra D-brane is required 
also because of 
the consistency condition associated with 
the ${\bf C}^2/{\bf Z}_M\vev{\sigma}$ singularity. 
The Ramond--Ramond tadpole cancellation condition \cite{RR} 
is given by \cite{RR6}
\begin{equation}
\tr \left( \gamma_{7;\sigma^k} \right) 
- 4 \sin^2\left( \frac{\pi k}{M} \right) 
  \tr \left( \gamma_{3;\sigma^k} \right) = 0, 
\qquad (k=1,\ldots,M-1).
\label{eq:RR6}
\end{equation}
All fractional D3-branes and D7-branes, no matter where
they are in the ${\bf C}$ direction, contribute to the condition.
%
Since the Ramond--Ramond tadpoles do not cancel 
in the U(3)$_{\rm H} \times$ U(6)$_{D7}$ model, 
extra D-branes have to be introduced.
We assume that there is a D-brane configuration which 
is consistent with the tadpole condition, 
and is free from unwanted extra massless particles.
Finding an explicit configuration is left to further investigation.

Even though we do not specify the D-brane configuration completely, 
there is still something we can learn. 
The U(1)$_{\rm H}$ vector field has to be massless, and hence 
should not couple to the twisted-sector Ramond--Ramond field.
Thus, the origin of the Fayet--Iliopoulos F-term parameter 
in (\ref{eq:super}) 
is not the VEV of the twisted NS--NS sector fields, but something 
else\footnote{This issue is discussed later again.}.
It is also easy to see that the gauge coupling constants of 
SU($N$)$_{\rm H}$ and U(1)$_{\rm H}$ do not satisfy a relation 
$\alpha_{N{\rm H}} = \alpha_{\rm 1H}$ at the string scale.

Let us now turn our attention to the six D7-branes. 
They are wrapped on a holomorphic 4-cycle so that 
the ${\cal N}$ = 1 SUSY is preserved \cite{BBS,OOY}.
The cycle is denoted by $\Sigma_1$ (see Fig.~\ref{fig:3+6}).
 
Now, the flavour symmetry U(6) becomes dynamical.
An SU(5) subgroup of this U(6) symmetry is identified with 
the SU(5)$_{\rm GUT}$.
The U(6) (and hence SU(5)$_{\rm GUT}$) gauge coupling constant 
is given by 
\begin{equation}
\frac{1}{\alpha_{U(6)}} = 
\frac{1}{g_s}\frac{{\rm vol}(\Sigma_1)}{(2\pi \sqrt{\alpha'})^4}, 
\end{equation}
while 
\begin{equation}
 \frac{1}{\alpha_{N{\rm H}}} = \frac{1}{g_s}.
\end{equation}
Thus, the string coupling is determined by the 
value of $\alpha_{N {\rm H}}$ --- $g_s \sim 1/2 \mbox{--}2$.
A moderately large volume of the cycle $\Sigma_1$ 
\begin{equation}
\frac{{\rm vol}(\Sigma_1)}{(2\pi \sqrt{\alpha'})^4} 
\sim (10\mbox{--}100)
\end{equation}
leads to the relatively weak coupling of SU(5)$_{\rm GUT}$ in 
(\ref{eq:disparity}), or equivalently, to the approximate unification 
of the MSSM gauge coupling constants.
Note also that the string scale is given by 
\begin{equation}
 \frac{1}{\sqrt{\alpha}'}
 = \left(
     \pi g_s^2
     \frac{(2\pi\sqrt{\alpha}')^6}{{\rm vol}(CY_3)}     
   \right)^{\frac{1}{2}} 
   M_{\rm pl}
  \simeq 
   \sqrt{\pi \alpha_{\rm GUT} \alpha_{N{\rm H}}}
   \left(
     \frac{(2\pi\sqrt{\alpha}')^2}{{\rm vol}(CY_3)/{\rm vol}(\Sigma_1)}
   \right)^{\frac{1}{2}} \times 2.4 \times 10^{18} \GEV,  
\label{eq:string-scl}
\end{equation}
and hence can be sufficiently low when the total volume 
of the Calabi--Yau 3-fold is sufficiently large.
Typically, $({\rm vol}(CY_3)/{\rm vol}(\Sigma_1))/4\pi \alpha' 
\sim 100$ is necessary for $1/\sqrt{\alpha'} \sim 10^{17}$ GeV.

The moderately large volume required above guarantees that 
the supergravity provides a good description to some extent, i.e., 
the vacuum is not in a purely stringy regime.
We also see from above that the Kaluza--Klein scale 
is roughly of the order of the GUT scale. 
Thus, it is tempting to speculate that the origin of the GUT scale, 
and hence of the Fayet--Iliopoulos term in (\ref{eq:super}), 
has something to do with the Kaluza--Klein scale. 

If the 3-form fluxes are introduced for the moduli stabilization 
\cite{flux},  the background 3-form field strength is of the order of 
$\alpha'/R^3$, and the 2-form potential of the order of $\alpha'/R^2$ 
\cite{KST}, where $R$ is the typical length scale of 
the internal manifold. 
Then, a dimension-two quantity (for the Fayet--Iliopoulos parameter 
$v^2$) is obtained from the background 2-form potential 
by multiplying $1/\alpha'$, 
which is $1/\alpha' \times (\alpha'/R^2) \sim 1/R^2$. 
Thus, it may be that the B-field background for the moduli 
stabilization is also the origin of the Fayet--Iliopoulos term 
\cite{IWY}.
Although the above idea is too naive, and neither the moduli 
stabilization nor back reaction 
is considered, 
yet it is an interesting idea for the origin of the GUT scale.

The U(6) symmetry has to be reduced to SU(5) (and extra U(1)'s).
To this end, non-trivial background of U(1) $\subset$ SU(6) field
strength is introduced; U(1) is a subgroup of SU(6) that commutes 
with SU(5). 
Let the Chan--Paton bundles on the six D7-branes be denoted by 
\begin{equation}
 \lambda_{({\bf 5}\oplus 1)\times ({\bf 5}^* \oplus 1)} \rightarrow
 \left(
\begin{array}{c|c}
  E_{{\bf 5}{\bf 5}^*} & E_{{\bf 5}6^*} \\
  \hline
  E_{6{\bf 5}^*} & E_{66^*}
 \end{array}
  \right). 
\end{equation}
The vector bundles $E_{{\bf 5}6^*}$ and $E_{6{\bf 5}^*}$ 
are non-trivial\footnote{The vector bundles have to be trivial around 
the GUT-breaking sector, though.}, 
and the spectrum no longer respects the U(6) symmetry.
The background field strength has to satisfy 
the ``generalized Hitchin equations'' in \cite{HvMr} 
so that the D = 4 ${\cal N}$ = 1 SUSY is preserved.



The spectrum of the D7--D7 open strings preserves only 
${\cal N}$ = 1 SUSY, unless the local geometry around 
the $\Sigma_1$ hypersurface satisfies special properties.
We have an SU(5) vector multiplet of ${\cal N}$ = 1 SUSY, which 
is identified with the SU(5)$_{\rm GUT}$ vector multiplet.
There are two U(1) symmetries coming from the D7--D7 open strings 
at the classical level. 
But, they usually have triangle anomalies, which are cancelled by 
the (generalized) Green--Schwarz mechanism. 
In this case, the vector fields of the symmetries 
are not massless.

The number of massless ${\cal N}$ = 1 SUSY chiral multiplets 
in the SU(5)$_{\rm GUT}$-{\bf adj.} representation is given by 
$h^0(T\Sigma_1\oplus {\cal N}_{\Sigma_1}) =
h^0(T\Sigma_1 \oplus K_{\Sigma_1})$ \cite{BSV,Adj.no-IIB}, 
where $T\Sigma_1$, ${\cal N}_{\Sigma_1}$ and $K_{\Sigma_1}$ 
are tangent, normal and canonical bundles on $\Sigma_1$, respectively, 
and $h^0$ stands for the number of global holomorphic sections 
of the corresponding vector bundles.
Thus, the SU(5)$_{\rm GUT}$-{\bf adj.} chiral multiplets are 
usually absent in the low-energy spectrum, just as desired 
in the SU(5)$_{\rm GUT} \times$U($N$)$_{\rm H}$ model ($N = 2,3$).
For the massless chiral multiplets from the open strings 
connecting five and the other D7-branes, see subsection 
\ref{ssec:Higgs}.

The Calabi-Yau geometry has been required so far to satisfy 
the following properties. It has to have a holomorphic 4-cycle, 
$\Sigma_1$, on which there is a point whose local geometry 
should be (approximately) ${\bf C}^2/{\bf Z}_M \times {\bf C}$.
The volume of the Calabi--Yau 3-fold 
is moderately large in units of string length 
in the directions both tangential and transverse 
to the cycle $\Sigma_1$. 
The cycle has to satisfy $h^0(T\Sigma_1 \oplus K_{\Sigma_1}) = 0$.
It is not hard to find a Calabi-Yau 3-fold that possesses 
the properties described above; one can find an example 
${\bf T}^6/{\bf Z}_{12}$ in \cite{WY01,WY02}.  
 
\subsection{\label{ssec:Higgs}Higgs Multiplets}

The SU(5)$_{\rm GUT} \times$ U(3)$_{\rm H}$ model requires 
Higgs multiplets, 
$H^i({\bf 5},{\bf 1}) + \bar{H}_i({\bf 5}^*,{\bf 1})$, in the spectrum.
It is economical if they 
are also obtained from the D7-branes wrapped on $\Sigma_1$.
Moreover, it is desirable to obtain the Higgs multiplets in this way, 
because they have non-vanishing wave functions 
at the ${\bf C}^2/{\bf Z}_M$ singularity 
(i.e., the locus of the GUT-breaking sector), and hence 
the interactions in the fourth line of (\ref{eq:super}) 
are expected not to be suppressed\footnote{
We do not discuss how the superpotential 
of the low-energy D = 4 effective theory is generated.}. 
Thus, we describe in this subsection how to obtain 
the Higgs multiplets from the open strings connecting 
five D7-branes and another(others) 
wrapping the same 4-cycle $\Sigma_1$.  

Since one D7-brane has already been introduced in addition 
to the five D7-branes for SU(5)$_{\rm GUT}$, let us first 
discuss whether it is possible to obtain both Higgs multiplets 
$H^i({\bf 5})$ and $\bar{H}({\bf 5}^*)$ from those six D7-branes.
The massless Higgs multiplets are in the spectrum when 
$h^0(E_{{\bf 5}6^*}\otimes (T\Sigma_1 \oplus K_{\Sigma_1})) = 1$ and 
$h^0(E_{6{\bf 5}^*}\otimes (T\Sigma_1 \oplus K_{\Sigma_1})) = 1$. 
Then, it follows that there is a global section also 
in the tensor product of the two vector bundles above.
This implies that 
\begin{equation}
h^0(\otimes^2 (T\Sigma_1 \oplus K_{\Sigma_1})) \geq 1, 
\label{eq:extra}
\label{eq:Higgs-from-6}
\end{equation}
since the vector bundle $E_{{\bf 5}6^*} \otimes E_{6{\bf 5}^*}$ 
is trivial.
Thus, we see that the geometry along the 4-cycle $\Sigma_1$ 
has to satisfy the extra condition (\ref{eq:extra}) 
in order that the two Higgs 
quintets are obtained from the six D7-branes.

There is another possibility where the two Higgs multiplets 
are obtained even when the 4-cycle $\Sigma_1$ does not satisfy 
the condition (\ref{eq:Higgs-from-6}). 
To this end, another D7-brane is introduced
which is also wrapped on the 4-cycle $\Sigma_1$. 
Let the Chan--Paton bundles for massless open strings be denoted by 
\begin{equation}
 \lambda_{({\bf 5}\oplus 1 \oplus 1) \times 
          ({\bf 5}\oplus 1 \oplus 1)^*} \rightarrow 
 \left(\begin{array}{c|c|c}
  E_{{\bf 5}{\bf 5}^*} & E_{{\bf 5}6^*}& E_{{\bf 5}7^*} \\
  \hline 
  E_{6{\bf 5}^*} & E_{66^*} & E_{67^*} \\
  \hline 
  E_{7{\bf 5}^*} & E_{76^*} & E_{77^*}
       \end{array}\right).
\end{equation}
The condition (\ref{eq:Higgs-from-6}) needs not be satisfied now, 
since, say, $E_{{\bf 5}7^*} \otimes E_{6{\bf 5}^*} \simeq E_{67^*}$ 
is not trivial generically.

The NMSSM is quite natural in this framework.
Suppose that 
$h^0(E_{6{\bf 5}^*}\otimes (T\Sigma_1 \oplus K_{\Sigma_1})) = 1$  
and $h^0(E_{{\bf 5}7^*}\otimes (T\Sigma_1 \oplus K_{\Sigma_1})) = 1$, 
so that the two Higgs quintets are obtained.
Then, it follows\footnote{$h^0(E_{6{\bf 5}^*}\otimes E_{{\bf 5}7^*}
\otimes \wedge^2 (T\Sigma_1 \oplus K_{\Sigma_1})) = 1$ does not 
follow when $E_{6{\bf 5}^*} \simeq E_{{\bf 5}7^*}$. 
However, we do not consider this case in this article. 
This is because the isomorphism implies that 
the Chan--Paton bundle $E_{{\bf 5}7^*}$ is also trivial 
around the ${\bf C}^2/{\bf Z}_M$ singularity 
in subsection \ref{ssec:GUT}, 
U(7) symmetry is enhanced there, and 
another massless hypermultiplet 
in the U(3)$_{\rm H}$-{\bf (anti)fund.} representation 
appears in the spectrum, invalidating the gauge-coupling unification.} 
that $h^0(E_{67^*}\otimes \wedge^2 (T\Sigma_1 \oplus K_{\Sigma_1})) 
\geq 1$. 
If this is due to a trivial bundle in 
$E_{67^*}\otimes \wedge^2 (T\Sigma_1 \oplus K_{\Sigma_1})$, 
then its dual bundle 
$E_{76^*} \otimes (T\Sigma_1 \oplus K_{\Sigma_1})$ also has a global 
holomorphic section.
Thus, another massless chiral multiplet is in the spectrum, 
which comes from the open strings connecting sixth and seventh 
D7-branes.
This chiral multiplet is a singlet of SU(5)$_{\rm GUT}$, which 
is denoted by $S$.  The interactions on the D7-branes predict 
a tri-linear coupling in the low-energy effective superpotential, 
\begin{equation}
 W \ni S^7_{\; 6} \bar{H}^6_{\; i} H^i_{\; 7},
\label{eq:NMSSM}
\end{equation}
which is nothing but the interaction of NMSSM \cite{NMSSM}.
The coefficient of this operator is of the order of 
the gauge coupling constant of SU(5)$_{\rm GUT}$ 
at the classical level. 
But, 
we do not understand all the dynamics (including
non-perturbative one) that generates necessary superpotential 
couplings, and hence it is impossible to derive a quantitative 
prediction for the value of this coupling constant.

What is shown above is a generalization of what was found 
in \cite{WY02}. There, the two Higgs quintets were obtained 
along with the singlet $S$ in the explicit toroidal orientifold 
model based on ${\bf T}^6/{\bf Z}_{12}$. 

Notice that the existence of the singlet $S$ is due 
to the property ${\cal N}_{\Sigma_1} \simeq K_{\Sigma_1}$,  
which is an immediate consequence of the definition of 
Calabi--Yau manifolds.
The NMSSM interaction (\ref{eq:NMSSM}) is the immediate 
consequence of the ${\cal N}$ = 4 SUSY interactions on the D7-branes.
The tri-linear interaction reflects the fact that 
the internal manifold has three complex dimensions.
Thus, it is quite interesting if the NMSSM interaction 
is discovered in future collider experiments.

\subsection{\label{ssec:QL}Quarks and Leptons}

Let us now illustrate the idea of how the quarks and leptons 
can be constructed in Type IIB vacua 
in a way consistent with the model 
we have constructed in the preceding subsections.
\begin{figure}[t]
 \begin{center}
\begin{picture}(200,200)(0,0)
   \resizebox{5cm}{!}{\includegraphics{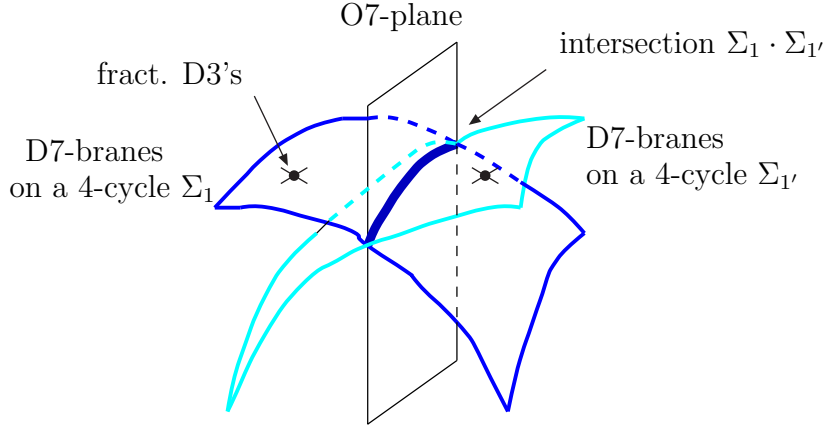}} 

   \Text(-70,150)[cb]{O7-plane}
   \Text(-160,105)[r]{D7-branes}
   \Text(-140,90)[r]{on a 4-cycle $\Sigma_1$}
   \Text(0,110)[l]{D7-branes}
   \Text(0,95)[l]{on a 4-cycle $\Sigma_{1'}$}

   \Line(-115,98)(-105,92)
   \Line(-115,92)(-105,98)
   \Vertex(-110,95){2}

   \Line(-43,98)(-33,92)
   \Line(-43,92)(-33,98)
   \Vertex(-38,95){2}

   \Text(-10,140)[lb]{intersection $\Sigma_1 \cdot \Sigma_{1'}$}
   \LongArrow(-15,135)(-43,115)

   \Text(-130,130)[rb]{fract. D3's}
   \LongArrow(-125,125)(-113,100)

\end{picture}
 \caption{A schematic picture of the configuration for 
 the chiral multiplets in the anti-symmetric-tensor representation. 
 A holomorphic 4-cycle $\Sigma_{1'}$ is the image of $\Sigma_1$ 
 under an involution associated with the O7-plane.
 They intersect at a holomorphic curve $\Sigma_1 \cdot \Sigma_{1'}$ 
 (thick curve in the figure).
 D7-branes are wrapped on the 4-cycles, and the 
 SU(5)$_{\rm GUT}$ gauge field propagates all over the world volume.
 The matter in the anti-symmetric-tensor representation arises at the 
 intersection. The GUT-breaking sector is realized by fractional
  D3-branes located at a ${\bf C}^2/{\bf Z}_2 \times {\bf C}$
  singularity on the 4-cycle $\Sigma_1$.}
\label{fig:asymrepr}
 \end{center}
\end{figure}
The model of unified theories we have considered is based on 
SU(5)-symmetric matter contents, and not on SO(10)-symmetric ones.
This is the case when the five D7-branes 
are not wrapped on the same cycle 
as the O7-plane and their orientifold images are (see 
Fig.~\ref{fig:asymrepr}).

On the other hand, a matter in the SU(5) anti-symmetric-tensor
representation arises on a locus where D-branes and their 
orientifold images coincide. Therefore, the D7-branes for 
the SU(5)$_{\rm GUT}$ gauge group should intersect 
with their orientifold mirror images in the Calabi--Yau 3-fold, and 
the anti-symmetric-tensor representation should arise there 
(see Fig.~\ref{fig:asymrepr}). 
SU(5)$_{\rm GUT}$-{\bf 10} representation is localized 
on a complex curve in the Calabi--Yau 3-fold (Fig.~\ref{fig:asymrepr}).

\subsubsection{\label{sssec:toy}Toy Model for Chiral Matter at D7--D7 
Intersection on Orbifold}

Let us first describe a simple toy model that shows the 
essential feature of how the chiral matter 
in the SU(5)$_{\rm GUT}$-{\bf 10} representation is obtained.  
An orbifold is used to construct the toy model, where 
the complex curve where the chiral matter lives is compact, 
while the total Calabi--Yau space is non-compact.
This semi-local model is considered as a model of 
the local neighbourhood of the curve where the D7-branes intersect.
We describe more generalized model in subsubsection \ref{sssec:curve}, 
where we do not restrict ourselves to orbifolds, 
in which arbitrary number of families of chiral matter 
can be obtained.

\begin{figure}[t]
 \begin{center}
\begin{picture}(250,250)(0,0)
   \resizebox{5cm}{!}{\includegraphics{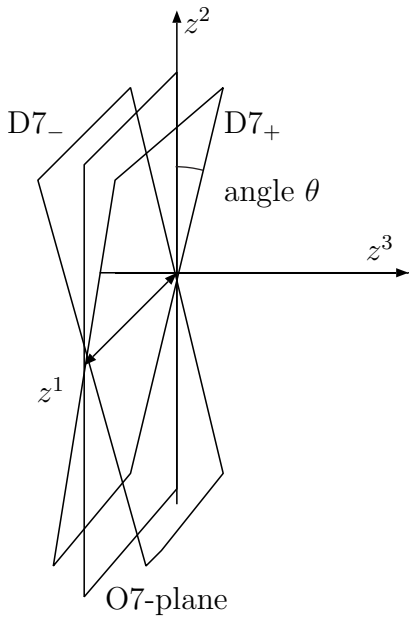}} 

   \Text(-11,134)[c]{$z^3$}
   \Text(-80,220)[c]{$z^2$}
   \Text(-130,85)[rt]{$z^1$}

   \CArc(-88,124)(40,75,90)
   \Text(-70,160)[lt]{angle $\theta$}
  
   \Text(-70,180)[l]{D7$_+$}
   \Text(-130,180)[r]{D7$_-$}
   \Text(-115,0)[l]{O7-plane}
\end{picture}
 \caption{Toy model explained in subsubsection \ref{sssec:toy}.
$z^1$ is the coordinate of ${\bf T}^2$, and $(z^2,z^3)$ are those 
of ${\bf C}^2$. Orientifold projection is associated with 
the reflection in $z^3$-direction, and hence the 
O7-plane is the fixed locus of the reflection, i.e., $z^3 = 0$ 
hypersurface.
D7$_-$ is the orientifold mirror image of D7$_+$.
  }
\label{fig:toy10}
 \end{center}
\end{figure}

Let us suppose that the six internal dimensions are of the form
${\bf T}^2 \times {\bf C}^2$, whose complex coordinates are 
$z_1$ for $T^2$ and $(z_2,z_3)$ for ${\bf C}^2$. 
Two stacks of space-filling D7-branes are stretched 
in $z_1 \wedge \bar{z}_1 \wedge 
( \cos \theta z_2 \pm \sin \theta z_3) \wedge 
( \cos \theta \bar{z}_2 \pm \sin \theta \bar{z}_3)$ 
(see Fig.~\ref{fig:toy10}). 
Here, $\theta$ is a constant angle and is arbitrary, 
unless it is an integral multiple of $\pi/2$. 
These stacks are mapped with each other by an SU(2)
transformation\footnote{Indeed, an SO(2) transformation between 
$z_2$ and $z_3$ by angle $\mp 2\theta$ does the job.}. 
Thus, we have a D = 6 (1,0)-SUSY gauge theory 
if the volume of $T^2$ is infinite, and a D = 4 
${\cal N}$ = 2 SUSY gauge theory if the volume is finite.
When the O7-plane is stretched in the $(z_1,z_2)$ complex planes, 
i.e., it is wrapped on the $z_3 = 0$ hypersurface, 
one stack of D7-branes D7$_-$ in Fig.~\ref{fig:toy10} 
is the orientifold image of the other stack D7$_+$.
The gauge theory consists only of a hypermultiplet 
in the anti-symmetric-tensor representation.

The system described above still preserves ${\cal N}$ = 2 SUSY, 
and the matter contents are vector-like in the D = 4 
effective theory. Thus, we impose an orbifold projection 
to obtain a chiral theory.
Let us consider a ${\bf Z}_N$ orbifold\footnote{The ${\bf Z}_N$ 
has to be a symmetry of $T^2$, and  $\alpha \in (2\pi / N) {\bf Z}$.}, 
where the three complex coordinates transform as 
\begin{equation}
 z_1 \mapsto e^{2i\alpha} z_1, \quad z_2 \mapsto e^{-i\alpha}z_2, 
 \quad z_3 \mapsto e^{-i\alpha}z_3.
\label{eq:toy-rot}
\end{equation}
This transformation belongs to SU(3), and hence the D = 4 
${\cal N}$ = 1 SUSY is preserved. 
Note that it also preserves the world volumes of 
both stacks of D7-branes, and hence extra D-branes do not have to be 
introduced as images of the ${\bf Z}_N$ action. 
When the matrix twisting the Chan--Paton indices is chosen suitably, 
two states among four in the NS sector and 
two states among four in the R sector 
survive the orbifold projection.
Thus, one chiral multiplet in the anti-symmetric-tensor representation is in the massless spectrum, while its hermitian conjugate 
is projected out.
For more details, see the appendix \ref{sec:toy10}.

\subsubsection{\label{sssec:curve}Model of Chiral Matter at 
D7--D7 Intersection Using Curved Manifold}

The toy model above yields an ${\cal N}$ = 1 chiral multiplet 
in the anti-symmetric-tensor representation, yet 
we have only one family; 
one hypermultiplet is obtained on ${\bf R}^{3,1} \times {\bf T}^2$, 
half of the massless modes (one chiral multiplet) 
remains in the spectrum, while the other half is projected out.
The only one family, however, is not a generic feature of the chiral
matters obtained at the intersection of D7-branes, 
if we do not restrict ourselves to the simplest model above, 
which uses ${\bf T}^2$ and its orbifolds. 

Suppose that there is an O7-plane in a Calabi--Yau 3-fold. 
A stack of D7-branes are wrapped on the holomorphic 4-cycle 
$\Sigma_1$ and another stack of D7-branes are on another 
holomorphic 4-cycle $\Sigma_{1'}$.
The 4-cycle $\Sigma_{1'}$ is supposed to be the image of $\Sigma_1$ 
under the involution associated with the orientifold plane.
We consider that each stack consists of six or seven D7-branes, 
or more, if necessary.
Yang--Mills fields on the D7-branes on $\Sigma_{1'}$ are identified 
with those on $\Sigma_1$. Their Kaluza--Klein zero modes   
have been discussed in subsection \ref{ssec:GUT} and \ref{ssec:Higgs}.
Now let us assume that the two 4-cycles $\Sigma_1$ and $\Sigma_{1'}$ 
intersect at a holomorphic curve $\Sigma_1 \cdot \Sigma_{1'}$ 
as in the toy model of subsubsection \ref{sssec:toy} 
(see Fig.~\ref{fig:asymrepr}). 
The massless matter contents localized at the intersection 
consist of two complex bosons and a Weyl fermion of D = 6 spacetime.
The two bosons are sections of 
$E_{\Sigma_1} \otimes E^*_{\Sigma_{1'}} \otimes 
({\cal N}_{\Sigma_1 \cdot \Sigma_{1'}|\Sigma_1} \otimes 
{\cal N}_{\Sigma_1 \cdot \Sigma_{1'}|\Sigma_{1'}})^{1/2}$ and 
$E^*_{\Sigma_1} \otimes E_{\Sigma_{1'}} \otimes 
({\cal N}_{\Sigma_1 \cdot \Sigma_{1'}|\Sigma_1} \otimes 
{\cal N}_{\Sigma_1 \cdot \Sigma_{1'} | \Sigma_{1'}})^{1/2}$ 
\cite{BSV,Iflow-curved,MM}, and the fermion is a section of 
$E_{\Sigma_1} \otimes E^*_{\Sigma_{1'}}$ 
\cite{Witten5,Iflow-curved,MM}; 
here, ${\cal N}_{\Sigma_1 \cdot \Sigma_{1'} |\Sigma_1}$ 
is the normal bundle on the intersection $\Sigma_1 \cdot \Sigma_{1'}$
associated with the embedding 
$(i_{\Sigma_{1'}})|_{\Sigma_1 \cdot \Sigma_{1'} } : 
\Sigma_1 \cdot \Sigma_{1'} \hookrightarrow \Sigma_1$, 
and ${\cal N}_{\Sigma_1 \cdot \Sigma_{1'} | \Sigma_{1'}}$ 
with the embedding 
$(i_{\Sigma_1})|_{\Sigma_1 \cdot \Sigma_{1'} } : 
\Sigma_1 \cdot \Sigma_{1'} \hookrightarrow \Sigma_{1'}$.
$E_{\Sigma_1}$ denotes the Chan--Paton U(1) bundle on $\Sigma_1$ 
that leads to 
$E_{\Sigma_1} \otimes E^*_{\Sigma_1} \simeq E_{{\bf 5}{\bf 5}^*} 
\oplus E_{{\bf 5}6^*} \oplus E_{6{\bf 5}^*} \oplus E_{66^*} 
\oplus \cdots$ introduced in subsection \ref{ssec:GUT} 
(and \ref{ssec:Higgs}).  
$E_{\Sigma_{1'}}$ plays the same role on $\Sigma_{1'}$ 
as $E_{\Sigma_1}$ does on $\Sigma_1$.

The complex curve $\Sigma_1 \cdot \Sigma_{1'}$ is compact, 
and hence we obtain massless modes of D = 4 theories through 
the Kaluza--Klein reduction of the two complex bosons and 
the Weyl fermion on the curve. 
The net number of massless complex scalar fields 
in chiral multiplets is given by 
\begin{eqnarray}
 N_B & = & 
     h^0(\Sigma_1 \cdot \Sigma_{1'},
             E_{\Sigma_1} \otimes E^*_{\Sigma_{1'}} \otimes 
             ({\cal N}_{\Sigma_1 \cdot \Sigma_{1'}|\Sigma_1} \otimes
              {\cal N}_{\Sigma_1 \cdot \Sigma_{1'}|\Sigma_{1'}})^{1/2}
          ) \nonumber \\ & & 
  -  h^0(\Sigma_1 \cdot \Sigma_{1'},
             E^*_{\Sigma_1} \otimes E_{\Sigma_{1'}} \otimes 
             ({\cal N}_{\Sigma_1 \cdot \Sigma_{1'}|\Sigma_1} \otimes
             {\cal N}_{\Sigma_1 \cdot \Sigma_{1'} | \Sigma_{1'}})^{1/2}
         )  \nonumber \\
     & = & 
     h^0(\Sigma_1 \cdot \Sigma_{1'},
             E_{\Sigma_1} \otimes E^*_{\Sigma_{1'}} \otimes 
             T^*(\Sigma_1 \cdot \Sigma_{1'})^{1/2}
         ) 
   - h^1(\Sigma_1 \cdot \Sigma_{1'},
             E_{\Sigma_1} \otimes E^*_{\Sigma_{1'}} \otimes 
             T^*(\Sigma_1 \cdot \Sigma_{1'})^{1/2}
         )  \nonumber \\
     & = & 
     \chi(\Sigma_1 \cdot \Sigma_{1'}, 
             E_{\Sigma_1} \otimes E^*_{\Sigma_{1'}} \otimes 
             T^*(\Sigma_1 \cdot \Sigma_{1'})^{1/2}
          ) \nonumber \\
     & = & \int_{\Sigma_1 \cdot \Sigma_{1'}} 
           {\rm ch}(E_{\Sigma_1} \otimes E^*_{\Sigma_{1'}})  
           e^{-\frac{1}{2}c_1(T(\Sigma_1 \cdot \Sigma_{1'}))}
           {\rm Td}(T(\Sigma_1 \cdot \Sigma_{1'})) 
    = \int_{\Sigma_1 \cdot \Sigma_{1'}} 
     \left(\frac{F_{\Sigma_1}}{2\pi}-\frac{F_{\Sigma_{1'}}}{2\pi}
     \right), 
\label{eq:NB}
\end{eqnarray}
where the Calabi--Yau condition 
$T^*(\Sigma_1 \cdot \Sigma_{1'}) \simeq  
{\cal N}_{\Sigma_1 \cdot \Sigma_{1'}|\Sigma_{1'}} \otimes
{\cal N}_{\Sigma_1 \cdot \Sigma_{1'}|\Sigma_1}$ and 
the Serre duality are used in the second equality, and 
the last equality is the Hirzebruch--Riemann--Roch 
formula \cite{Hirzebruch}. 
Here, $\chi$, ${\rm ch}$, and ${\rm Td}$ are 
the Euler characteristic, Chern character, and Todd class 
of complex vector bundles\footnote{For their 
definitions, see \cite{Hirzebruch}.}.
The number of chiral fermions is given by 
\begin{eqnarray}
 N_F & = & {\rm index}_{\Sigma_1 \cdot \Sigma_{1'}} 
    \Dsl_{ \left(E_{\Sigma_1} \otimes E^*_{\Sigma_{1'}} 
           \right)
          }  \nonumber \\
     & = & 
     \int_{\Sigma_1 \cdot \Sigma_{1'}} 
        {\rm ch}(E_{\Sigma_1} \otimes E^*_{\Sigma_{1'}} 
                )
         \hat{A}(T(\Sigma_1 \cdot \Sigma_{1'}))  \label{eq:NF} \\
     & = & 
     \int_{\Sigma_1 \cdot \Sigma_{1'}} 
           {\rm ch}(E_{\Sigma_1} \otimes E^*_{\Sigma_{1'}} 
            ) e^{-\frac{1}{2}c_1(T(\Sigma_1 \cdot \Sigma_{1'}))}
           {\rm Td}(T(\Sigma_1 \cdot \Sigma_{1'})) 
   = \int_{\Sigma_1 \cdot \Sigma_{1'}} 
    \left( \frac{F_{\Sigma_1}}{2\pi}- \frac{F_{\Sigma_{1'}}}{2\pi}
    \right),    \nonumber  
\end{eqnarray}
where the first line stands for the index of the Dirac operator 
on the curve $\Sigma_1 \cdot \Sigma_{1'}$, and $\hat{A}$ is the 
$\hat{A}$ class\footnote{See \cite{Hirzebruch} for definition.}. 
Therefore, the number of massless chiral bosonic modes agrees with 
that of fermionic modes, as expected from D = 4 ${\cal N}$ = 1 SUSY, 
and $N_B = N_F$ gives the number of families of chiral 
multiplets for each irreducible sub-bundle (sub-representation)  
of the Chan--Paton bundle 
$E_{\Sigma_1} \otimes E^*_{\Sigma_{1'}}$.
Note that all of $E_{\Sigma_1}$ and $E_{\Sigma_{1'}}$ in 
(\ref{eq:NB}) and (\ref{eq:NF}) should be, 
precisely speaking, replaced by 
$((i_{\Sigma_{1'}})|_{ \Sigma_1 \cdot \Sigma_{1'} })^* E_{\Sigma_1}$ 
and 
$((i_{\Sigma_1})|_{ \Sigma_1 \cdot \Sigma_{1'} })^* E_{\Sigma_{1'}}$, 
respectively; the abuse of notations is just 
for visual clarity.
The formulae (\ref{eq:NB}) and (\ref{eq:NF}) are nothing but 
a local version of the formula of the number of chiral multiplets 
in \cite{Dquint}, where it is given by paring of the Ramond--Ramond 
charges of the D-branes \cite{Mukai,Iflow,HvMr,Iflow-curved,MM,MSSSS}. 
For the relation between the formula in \cite{Dquint} and 
(\ref{eq:NB},\ref{eq:NF}), see the appendix \ref{sec:FamilyK}.

The formula of the number of chiral multiplets 
(\ref{eq:NB},\ref{eq:NF}), or equivalently, 
(\ref{eq:pairing}) in the appendix \ref{sec:FamilyK}, 
is applied to two stacks of ordinary D-branes. 
The multiplicity\footnote{New material added in version 2, March, 
2006.} of chiral multiplets in the anti-symmetric 
representation that arise at the intersection of a stack of 
D7-branes and an O7-plane (and the mirror-image stack of D7-branes)
\cite{Roemelsberger,Berlin} is given in local form 
\cite{Oplane,Toronto} by (see also the appendix B)
\begin{equation}
\frac{1}{2} \left[ \int_{\Sigma_1 \cdot \Sigma_{1'}} 
  \left( \frac{F_{\Sigma_1}}{2\pi}-\frac{F_{\Sigma_{1'}}}{2\pi}\right) 
- \int_{\Sigma_1 \cdot W} 
  2 \left(\frac{F_{\Sigma_1}}{2\pi}- \frac{B}{(2\pi)^2\alpha'}\right)
\right] = \int_{\Sigma_1 \cdot W} 
  2 \left(\frac{F_{\Sigma_1}}{2\pi}- \frac{B}{(2\pi)^2\alpha'}\right).
\label{eq:Nasym}
\end{equation}
Here, $W$ is the locus of an O7-plane. This can be an odd integer 
such as 3, since the $B$ field on a curve on an O7-plane is either 
integral or half integral \cite{Oplane,FrWt}.

\begin{figure}[t]
 \begin{center}
\begin{picture}(250,250)(0,0)
   \resizebox{8cm}{!}{\includegraphics{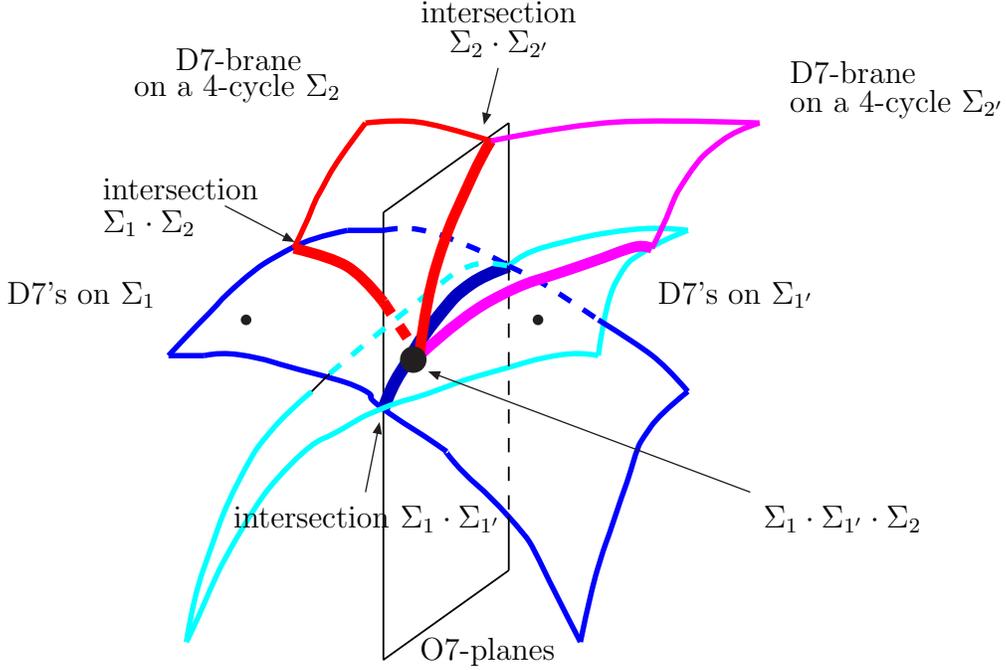}} 


   \Vertex(-195,130){2}
   \Vertex(-85,130){2}

   \Text(-130,10)[lt]{O7-planes}

   \Text(-230,140)[r]{D7's on $\Sigma_1$}
   \Text(-40,140)[l]{D7's on $\Sigma_{1'}$}

   \Text(-175,225)[rb]{D7-brane}
   \Text(-160,213)[rb]{on a 4-cycle $\Sigma_2$}
   \Text(10,220)[lb]{D7-brane}
   \Text(10,207)[lb]{on a 4-cycle $\Sigma_{2'}$}

   \Text(-100,243)[cb]{intersection}
   \Text(-100,230)[cb]{$\Sigma_2 \cdot \Sigma_{2'}$}
   \LongArrow(-100,225)(-105,205)

   \Text(-250,180)[lc]{intersection}
   \Text(-250,167)[lc]{$\Sigma_1 \cdot \Sigma_2$}
   \LongArrow(-203,173)(-178,160)

   \Text(-150,60)[ct]{intersection $\Sigma_1 \cdot \Sigma_{1'}$}
   \LongArrow(-150,65)(-145,90)

   \Vertex(-132,115){5}
   \Text(0,60)[lt]{$\Sigma_1 \cdot \Sigma_{1'} \cdot \Sigma_2$}
   \LongArrow(-5,65)(-125,110)
\end{picture}
 \caption{A schematic picture of the D-brane configuration for  
quarks and leptons. The SU(5)$_{\rm GUT}$ gauge field 
propagates on the D7-branes on $\Sigma_1$ and $\Sigma_{1'}$.
The chiral multiplets in the {\bf 10} representation are localized 
on $\Sigma_1 \cdot \Sigma_{1'}$, and 
those in the ${\bf 5}^*$ representation (and right-handed neutrinos) 
on $\Sigma_1 \cdot \Sigma_2$. The GUT-breaking sector is denoted by 
a small dot on $\Sigma_1$ (and its orientifold 
image on $\Sigma_{1'}$).
} 
\label{fig:allcast}
 \end{center}
\end{figure}

Let us turn our attention to the chiral multiplets in the 
{\bf 5}$^*$ representation. 
They are obtained just in the same way as the chiral multiplets 
in the {\bf 10} representation are obtained.
To be more explicit, another D7-brane wrapped on a holomorphic 
4-cycle $\Sigma_2$ is introduced, and $\Sigma_2$ is assumed 
to intersect with $\Sigma_1$ on a complex curve 
$\Sigma_1 \cdot \Sigma_2$ (see Fig.~\ref{fig:allcast}).
Massless particles in the $({\bf 5},{\bf 1}^*)$ 
(or $({\bf 5}^*,{\bf 1})$) representation arise at the intersection
from the open strings connecting the D7-branes 
on $\Sigma_1$ and $\Sigma_2$.
After the Kaluza--Klein reduction on the complex curve, 
D = 4 massless matter contents can be chiral 
when the Chan--Paton bundles 
on $\Sigma_1$ and $\Sigma_2$ are suitably chosen.
The number of massless chiral multiplets in the ${\bf 5}^*$ 
representation is given by (\ref{eq:NB},\ref{eq:NF}) with 
$\Sigma_{1'}$ replaced by $\Sigma_2$.

The mechanism to obtain chiral matter discussed so far 
is essentially the one in \cite{Dquint,Berlin}.

A U(1) symmetry arises from the D7-brane wrapped on $\Sigma_2$, 
which is denoted by U(1)$_0$.
Thus, the chiral multiplets obtained above is not 
in the {\bf 5}$^*$ representation, but in a bi-fundamental
representation $({\bf 5}^*,{\bf 1})$ 
under SU(5)$_{\rm GUT} \times$ U(1)$_0$.
However, this is not a problem.
The U(1)$_0$ symmetry has mixed anomaly with SU(5)$_{\rm GUT}$ 
because of the chiral multiplets. All the anomalous U(1) vector 
multiplets become massive through the Green--Schwarz interactions, 
and are irrelevant to the low-energy physics \cite{MadridU(1)}.
Likewise, the centre-of-mass U(1) symmetry arising from the 
first five D7-branes on $\Sigma_1$ is denoted by $U(1)_5$, and 
the U(1) symmetry from the sixth D7-brane on $\Sigma_1$ by U(1)$_6$ 
(and the one from the seventh D7-brane on $\Sigma_1$ by U(1)$_7$, 
if it exists). They also have  mixed anomaly with 
SU(5)$_{\rm GUT}$.
Some of linear combinations of these U(1) symmetries can be 
free from anomaly, and have corresponding massless gauge fields.
Let us consider the following linear combination, which is free from  
the mixed anomaly with SU(5)$_{\rm GUT}$:
\begin{equation}
 1 \times U(1)_5 + (-5) \times U(1)_0 + 5 \times U(1)_6  \; \; \;
(+ 5 \times U(1)_7 + \cdots).
\end{equation}
The {\bf 10} representation is charged by $+2$, ${\bf 5}^*$ by $-6$, 
$\bar{H}^6_{\; \; i} ({\bf 5}^*)$ by $+4$, and 
$H^i_{\; \; 7 {\rm ~or~} 6}({\bf 5})$ by $-4$ under this symmetry, 
and hence we identify this symmetry with the U(1)$_{B-L}$ symmetry.
The triangle anomaly of itself vanishes 
when there are three chiral multiplets  
whose U(1)$_{B-L}$ charge is $-10$. 
These multiplets are identified with right-handed neutrinos. 
Since the open strings connecting the D7-brane on $\Sigma_2$ and 
the sixth (or seventh) D7-brane on $\Sigma_1$ carries U(1)$_{B-L}$ 
charge by $\pm 10$, we expect the right-handed neutrinos arise 
from those strings.
Therefore, it follows that the chiral multiplets in the ${\bf 5}^*$ 
representation and the right-handed neutrinos are localized 
at the intersection $\Sigma_1 \cdot \Sigma_2$, whereas those 
in the {\bf 10} representation are at the intersection 
$\Sigma_1 \cdot \Sigma_{1'}$ (Fig.~\ref{fig:allcast}).

\subsubsection{\label{sssec:RR-family}Ramond--Ramond Charge 
Cancellation and Origin of Family Structure}

We have not discussed the cancellation of the Ramond--Ramond 
charges of D-branes. This subsubsection is devoted in describing, 
in a qualitative manner, how the charges can be cancelled.
We do not try to obtain an explicit D-brane configuration 
and background geometry in this article\footnote{Thus, it is not 
guaranteed that there is a consistent solution that realizes 
the idea described in this article. We just assume in this article 
that there is.}. Although some of 
phenomenological aspects depend on the explicit solutions, 
there are also some generic features that do not depend 
on such details.
One will see at the end of this subsubsection that 
the qualitative understanding of the Ramond--Ramond charge 
cancellation suggests us a possible geometric origin of family 
structure of quarks and leptons.

The Bianchi identities of the Ramond--Ramond potentials are given 
by \cite{BwB,Iflow,Iflow-curved,MM}\footnote{Contributions from 
the background fluxes are not taken into account here.}
\begin{equation}
 d G = \sum_k v_X(\Sigma_k,E_{\Sigma_k}), 
\label{eq:Bianchi}
\end{equation}
where $G$ is the sum of field strengths of the various 
Ramond--Ramond potentials and $v_X(\Sigma_k,E_{\Sigma_k})$ is 
the Ramond--Ramond charge of a D-brane wrapped on a cycle $\Sigma_k$. 
$E_{\Sigma_k}$ stands for the Chan--Paton bundle on it.
The explicit expression for the Ramond--Ramond charge 
\cite{Mukai,Iflow,HvMr,Iflow-curved,MM,MSSSS} 
is given in (\ref{eq:RRchargeA},\ref{eq:RRchargeB}) 
in the appendix \ref{sec:FamilyK} for convenience.
The both sides of (\ref{eq:Bianchi}) are in even-dimensional 
cohomology of the Calabi--Yau 3-fold. 
Now, the both sides are integrated on an even-dimensional cycle, and 
then, the Ramond--Ramond charge cancellation condition follows:
\begin{equation}
 \int_{{\rm compact~cycle~of~}X}\sum_k v_X(\Sigma_k,E_{\Sigma_k}) = 0.
\label{eq:RRcancel}
\end{equation}
We are not interested in the condition (\ref{eq:RRcancel}) for 
the total Calabi--Yau 3-fold (6-cycle), since we are concerned 
only about the local model.
Conditions only for compact 2-cycles and 4-cycles relevant to  
our construction are considered in the following.

The intersection of two holomorphic 4-cycles $\Sigma$ and $\Sigma'$ 
is a compact 2-cycle $\Sigma \cdot \Sigma'$, 
and hence we have a charge cancellation 
condition associated with this cycle. 
This type of condition is applied for 
$\Sigma_1 \cdot \Sigma_{1'}$, $\Sigma_1 \cdot \Sigma_2$, etc..
A D7-brane wrapped on $\Sigma'$ contributes to the condition for 
$\Sigma \cdot \Sigma'$ by 
\begin{equation}
 \int_{\Sigma \cdot \Sigma'}(i_{\Sigma'})_*(1) = 
\# (\Sigma \cdot \Sigma' \cdot \Sigma' ),
\end{equation}
i.e., by the intersection number of $\Sigma$, $\Sigma'$, and again, 
$\Sigma'$. $N$ D7-branes wrapped on $\Sigma$ contribute 
to the charge cancellation condition by 
$N \times \# (\Sigma \cdot \Sigma' \cdot \Sigma) $.
These contributions are, in general, non-zero, and have to be 
cancelled by other contributions from D7-branes 
wrapped on other 4-cycles. A D7-brane wrapped on a 4-cycle $\Sigma''$ 
contribute by the intersection number 
$\# (\Sigma \cdot \Sigma' \cdot \Sigma'')$.

A compact 4-cycle $\Sigma$ also has a condition, 
which is obtained by integrating the Bianchi identity 
(\ref{eq:Bianchi}) over it. 
This type of condition is applied for $\Sigma_1$ and $\Sigma_2$.
D7-branes wrapped on $\Sigma$ itself contribute 
by\footnote{The self-intersection formula for hypersurface 
$(i_\Sigma)^* (i_\Sigma)_* (1) = 
c_1({\cal N}_{\Sigma | X})$ is used.}\raisebox{1.5mm}{,}\footnote{The 
expression (\ref{eq:RRchargeB}) is used for the Ramond--Ramond charge 
of D-branes. See also footnote 17.}  
\begin{eqnarray}
 \int_{\Sigma} (i_{\Sigma})_* 
    \left({\rm ch}(E_{\Sigma} \otimes {\cal L}_B^{-1})
    \right)_{\rm 2-form} 
& = & \int_{\Sigma} c_1({\cal N}_{\Sigma | X}) \wedge 
    \left({\rm ch}(E_{\Sigma} \otimes {\cal L}_B^{-1})
    \right)_{\rm 2-form} \nonumber \\
& & = \int_{\Sigma \cdot \Sigma} 
  \left(\frac{F_{\Sigma}}{2\pi}-\frac{B}{(2\pi)^2\alpha'}\right).    
\end{eqnarray}
Other D7-branes (wrapped on a 4-cycle $\Sigma^{'''}$) 
contribute by 
\begin{eqnarray}
 \int_{\Sigma} (i_{\Sigma^{'''}})_* 
    \left({\rm ch}(E_{\Sigma^{'''}} \otimes {\cal L}_B^{-1})
    \right)_{\rm 2-form} 
 & = & \int_{\Sigma \cdot \Sigma^{'''}} 
    ((i_{\Sigma})|_{\Sigma \cdot \Sigma^{'''}})^* 
    \left({\rm ch}(E_{\Sigma^{'''}}\otimes {\cal L}_B^{-1})
    \right)_{\rm 2-form}  \nonumber \\
& &  = \int_{\Sigma \cdot \Sigma^{'''}}
  \left(\frac{F_{\Sigma^{'''}}}{2\pi}-\frac{B}{(2\pi)^2\alpha'}\right).
\end{eqnarray}
Here, ${\cal L}_B$ is a ``line bundle'' whose first Chern class is 
$B/((2\pi)^2\alpha')$.
D5-branes, including the fractional D3-branes, also contribute to 
the condition.
These contributions have to cancel one another for each 4-cycle.
Notice that the effects of the orientifold projection has not 
been considered seriously, and thus the above argument is 
only qualitative.

The Chern--Simons interaction on D-branes guarantees 
that the theory is locally anomaly free at any points of 
the D = 10 Type IIB string theory.
Localization of fermions and that of Ramond--Ramond charge 
are related through the Chern--Simons interaction, 
and the charge gives rise to proper inflow of anomaly 
for the localized fermion \cite{Iflow,Iflow-curved,MM,MSSSS}. 
Therefore, D = 4 theories obtained after compactification 
are free from anomalies, as long as the total sum of 
the anomaly inflow vanishes, or in other words, 
as long as the sum of Ramond--Ramond charges 
vanishes on every compact cycle.
The Ramond--Ramond charge cancellation on 4-cycles is responsible 
for the triangle anomalies on the D7 world volumes, 
and the cancellation on 2-cycles both for triangle anomalies on D5  
and box anomalies on D7 world volumes.

Incidentally, there have been proposed phenomenological models 
of the family structure of quarks and leptons that use 
anomaly inflow and cancellation in the internal space 
\cite{WYfamily,HbMr}. 
\begin{figure}[t]
 \begin{center}
\begin{picture}(250,150)(50,0)
   \resizebox{13cm}{!}{\includegraphics{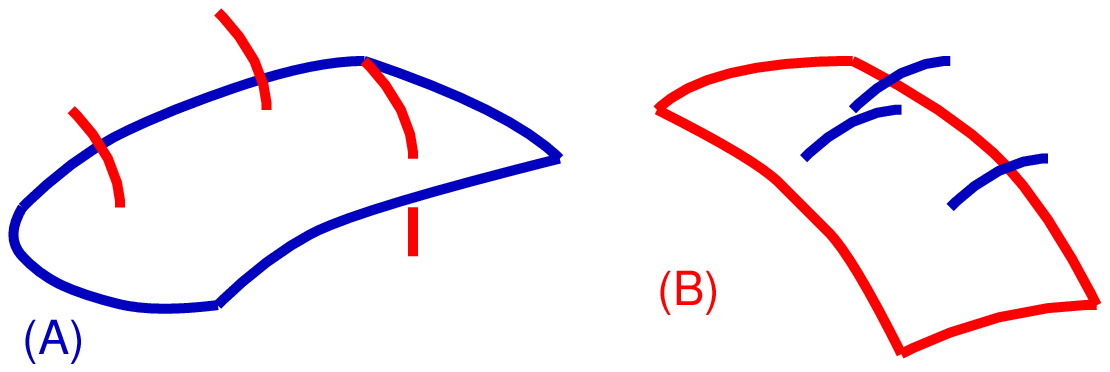}} 









   \Vertex(-328,56){3}
   \Vertex(-280,85){3}
   \Vertex(-230,70){3}
   \Text(-293,25)[lt]{$\Sigma_1 \cdot \Sigma_{1'}$({\bf 10})}
   \Text(-300,125)[cb]{$\Sigma_1 \cdot \Sigma_2$({\bf 5}$^*$+{\bf 1})}
   \LongArrow(-350,115)(-285,87)
   \Text(-355,115)[r]{$\Sigma_1 \cdot \Sigma_{1'} \cdot \Sigma_2$}

   \Vertex(-100,72){3}
   \Vertex(-83,85){3}
   \Vertex(-50,55){3}
   \Text(-75,15)[rt]{$\Sigma_1 \cdot \Sigma_2$
                      ({\bf 5}$^*$+{\bf 1})}
   \Text(-15,70)[l]{$\Sigma_1 \cdot \Sigma_{1'}$({\bf 10})}
   \LongArrow(-100,120)(-100,77)
   \Text(-100,123)[cb]{$\Sigma_1 \cdot \Sigma_{1'} \cdot \Sigma_2$}
%
\end{picture}
 \caption{Phenomenological models of family structure 
in \cite{WYfamily} can be considered as effective theories 
obtained by projecting onto the complex curve 
$\Sigma_1 \cdot \Sigma_{1'}$ (A). Multiplets in the {\bf 5}$^*$ 
representation and right-handed neutrinos 
are localized in the internal space, and 
the large mixing between them may be understood in terms of geometry.
Another model in \cite{HbMr} can be considered as an effective theory 
obtained by projecting onto the complex curve 
$\Sigma_1 \cdot \Sigma_2$ (B). 
Multiplets in the {\bf 10} representation 
are localized in the internal space, and the large hierarchy 
between them may be explained geometrically.  
} 
\label{fig:family}
 \end{center}
\end{figure}
It is one of the biggest mysteries in the context of unified theories 
why and how the mixing angles of the SU(2)$_L$ interaction are small 
in the quark sector while large in the lepton sector 
\cite{SKatms,SKsolar,SNO,KamLAND}. 
Models in \cite{WYfamily,HbMr} 
are phenomenological approaches to this mystery. 
Chiral multiplets in the SU(5)$_{\rm GUT}$-{\bf 10} 
representation and those in the SU(5)$_{\rm GUT}$-{\bf 5}$^*$ 
representation are assumed to have totally different wavefunctions 
(i.e., localization properties). 
Proposed there are ideas of geometric realization of the origin 
of non-parallel family structure \cite{SY,Ramond} 
desired phenomenologically.

It is a remarkable coincidence that the D-brane configuration 
obtained in this article happens to support the phenomenologically  
motivated models of the family structure 
\cite{WYfamily,HbMr};  
the models were proposed  totally independent 
from the D-brane realization of unified theories considered 
in this article.
The multiplets in the ${\bf 10}$ representation are localized 
on the intersection $\Sigma_1 \cdot \Sigma_{1'}$, 
while those in ${\bf 5}^*$ and the right-handed neutrinos are 
on $\Sigma_1 \cdot \Sigma_2$. 
Let us suppose that the hypersurfaces $\Sigma_1$, $\Sigma_{1'}$ 
and $\Sigma_2$ have non-vanishing intersection number, i.e., 
$\# (\Sigma_1 \cdot \Sigma_{1'} \cdot \Sigma_2) \neq 0$. 
Then, models in \cite{WYfamily} 
can be considered as the effective theories 
obtained by projecting the model in this article onto the 
complex curve $\Sigma_1 \cdot \Sigma_{1'}$ 
(Fig.~\ref{fig:family} (A)). 
There, the anomaly due to the chiral zero modes 
in the ${\bf 10}$ representation on the complex curve 
$\Sigma_1 \cdot \Sigma_{1'}$ flows into some points 
$\Sigma_1 \cdot \Sigma_{1'} \cdot \Sigma_2$, where the anomaly 
is cancelled by chiral zero modes in the {\bf 5}$^*$ representation 
and the right-handed neutrinos.  
The models in \cite{HbMr} can be considered as the effective theory 
obtained by projecting the model in this article onto 
the complex curve $\Sigma_1 \cdot \Sigma_2$ 
(Fig.~\ref{fig:family} (B)). 
There, the anomaly due to the chiral zero modes 
in ${\bf 5}^*$ and the right-handed neutrinos flows into some points 
$\Sigma_1 \cdot \Sigma_{1'} \cdot \Sigma_2$, where the anomaly 
is cancelled by chiral zero modes in the {\bf 10} representation.
Therefore, the framework in this subsection, in which ${\bf 10}$'s 
and ${\bf 5}^*$'s are obtained at (different) intersections 
of D7-branes,  
not only exhibits the origin of the non-parallel family structure, 
but also (hopefully) makes it possible to obtain better 
understanding of the origin of the family structure 
and Yukawa coupling of all the quarks and leptons.

\section{\label{sec:Summary}Summary and Open Questions}

The absence of the large mass for the two Higgs doublets and 
the suppressed dimensions-five proton decay give us important clues  
to how the unified theory would be. The product-group unification 
models, based on SU(5)$_{\rm GUT} \times$U($N$)$_{\rm H}$ gauge 
group $(N=2,3)$, are one of a few classes of theories in which 
the two clues above are understood in terms of a single symmetry.
A preceding work \cite{IWY} suggests that the brane-world 
picture may exist behind the models.   
Based on this motivation, we have illustrated an idea of how to embed 
the model into the Type IIB string theory.

The Type IIB string theory is compactified on an 
orientifolded Calabi--Yau 3-fold, and space-filling 
D-branes are wrapped on holomorphic cycles of the Calabi--Yau 3-fold, 
so that D = 4 ${\cal N}$ = 1 SUSY is preserved. 
Various particles of the model, including quarks and leptons, 
and particles in the GUT-symmetry-breaking sector, are 
obtained as the massless modes contained in open strings connecting 
those D-branes. 
We do not restrict ourselves to toroidal orbifolds as the candidates 
of the Calabi--Yau 3-fold. 
We allow ourselves to choose a generic curved Calabi--Yau manifold, 
so that we can hope to find a suitable manifold that accommodates 
all the particles we need.
Although one can no longer hope to calculate {\it everything} 
without a CFT formulation, it is certainly not our primary interest. 
Instead, local geometry and local D-brane configuration 
are determined so that phenomenologically desired particle contents 
are obtained. 
Once the local configuration is fixed, then it may be possible 
to derive some new phenomenological implications, if not 
predictions; local information is sufficient for limited purposes. 
This is what we hope to do. 

The GUT-breaking sector is realized by a D3--D7 system put on 
the local geometry ${\bf C}^2/{\bf Z}_M \times {\bf C}$. 
The U($N$)$_{\rm H}$ gauge group is realized by fractional D3-branes.
The SU(5)$_{\rm GUT}$ gauge group is realized by D7-branes wrapped 
on a 4-cycle $\Sigma_1$.
Phenomenology requires that 
${\rm vol}(\Sigma_1)/(2\pi\sqrt{\alpha'})^4 \sim 10\mbox{--}100$  
and $g_s \sim (1/2\mbox{--}2)$.
The total volume is determined by 
${\rm vol}(CY_3)/{\rm vol}(\Sigma_1) / (2\pi\sqrt{\alpha'})^2 \sim 100$, 
when the string scale $1/\sqrt{\alpha'}$ is of the order of 
$10^{17}$ GeV.
Although the origin of the GUT scale is not clarified in this article, 
it may be related to the Kaluza--Klein scale.

The SU(5)$_{\rm GUT} \times$ U(3)$_{\rm H}$ model \cite{IY} 
contains Higgs multiplets, 
which are fundamental and anti-fundamental representations 
of SU(5)$_{\rm GUT}$. 
This model realizes the doublet--triplet splitting through the 
missing partner mechanism, and hence, there should be sizable coupling 
between the Higgs multiplets and the particles in the
GUT-symmetry-breaking sector.
Therefore, it is both economical and phenomenologically desirable,  
if the Higgs particles are obtained from open strings 
connecting the D7-branes wrapped on $\Sigma_1$.
If a geometry around the 4-cycle $\Sigma_1$ satisfies a particular 
condition, then the two Higgs multiplets are obtained from six
D7-branes. If the condition is not satisfied, one can obtain them 
from seven D7-branes. 
The gauge singlet of the NMSSM and its tri-linear interaction with 
the two Higgs doublets can also be realized easily as a special case.

The chiral quarks and leptons are obtained at intersections of two
stacks of D7-branes. Chiral multiplets in the {\bf 10} representation 
are obtained at the intersection of five D7-branes and 
their orientifold mirror images. 
Those in the {\bf 5}$^*$ representation and the right-handed neutrinos 
are obtained at an intersection of the five D7-branes and 
another D7-brane.
D = 4 chiral theories are obtained if the Chan--Paton bundles 
on the D7-branes are suitably chosen. 
The number of family is given by the pairing of the Ramond--Ramond 
charges of the D-branes (and orientifold planes).

The multiplets in the {\bf 10} representation are localized in 
one D7--D7 intersection and those in the {\bf 5}$^*$ representation 
and the right-handed neutrinos are at another. 
One of the important consequences is that this configuration 
leads to a geometric realization of the non-parallel family 
structure, which was proposed to explain the large mixing angles 
in the neutrino oscillation.
The other important consequence is that 
the rate of the proton decay through dimension-six operators 
is enhanced, and the branching ratios are modified, 
as in \cite{G2-localB}.

In this article, we have discussed properties that local configuration 
has to satisfy, and have illustrated an idea of how to construct 
a model that satisfies those properties. 
But, this article does not present an explicit model where 
the Ramond--Ramond charges are cancelled at all the cycles 
relevant to the model. 
That is, the existence of the consistent solution is not guaranteed.
If one further pursues to find an explicit solution, 
one might be able to predict extra particles (e.g., such  
as those in \cite{KMY}) that can be observable in future detectors.
One might also be able to determine whether the singlet 
of the NMSSM really exists in the spectrum or not.
It may also be that 
the idea of lifting the D = 4  models to the Type IIB string 
theory is excluded because of the absence of a consistent solution.

The origin of Yukawa coupling and other interactions in the
superpotential is also poorly understood. 
Some of them are obtained through perturbative interactions of the Type
IIB theory, but not all of them. 
Although the discrete $R$ symmetry plays quite an important 
role in the D = 4 models, the origin of this symmetry 
is not identified, either. 
We leave these issues to further investigation.

\section*{Acknowledgements} 
The authors are grateful to the Theory Division of CERN 
for the hospitality, where earlier part of this work was done. 
T.W. thanks the Japan Society for the Promotion of Science 
and the Miller Institute for the Basic Research of Science. 
This work is partially supported by 
the Director, Office of Science, Office of High Energy and 
Nuclear Physics, of the U.S. Department of Energy 
under Contract DE-AC03-76SF00098 (T.W.), 
and Grant-in-Aid Scientific Research (s) 14102004 (T.Y.).

\appendix

\section{\label{sec:toy10}Toy Model of Chiral Anti-Symmetric-Tensor Representation at the Intersection of D7-branes}

This appendix provides details of the toy model 
sketched in subsubsection \ref{sssec:toy}.
One can see from the model how chiral matter in anti-symmetric
representation is obtained at the intersection of two stacks 
of D7-branes.

We first consider a background geometry 
${\bf R}^{3,1} \times {\bf T}^2 \times {\bf C}^2$. 
Coordinates $x^{0,1,2,3}$, $z^1$, and 
$(z^2 \equiv (x^6+ix^7),z^3 \equiv (x^8+ix^9))$ are used 
for ${\bf R}^{3,1}$, ${\bf T}^2$ and ${\bf C}^2$, respectively.
The ``internal'' space is ${\bf T}^2 \times {\bf C}^2$ and 
is non-compact. 
The non-compact geometry is regarded as a local geometry 
around the intersection of D7-branes we are interested in.
Two stacks of ${\bf R}^{3,1}$-filling D7-branes are introduced.
One of them, which we refer to as D7$_-$, is stretched 
in $z^1 \wedge \bar{z}^1 \wedge 
(\cos \theta z^2 - \sin \theta z^3) \wedge 
(\cos \theta \bar{z}^2 - \sin \theta \bar{z}^3)$, 
and the other, which we refer to as D7$_+$,  
in  $z^1 \wedge \bar{z}^1 \wedge 
(\cos \theta z^2 + \sin \theta z^3) \wedge 
(\cos \theta \bar{z}^2 + \sin \theta \bar{z}^3)$.
They intersect at a complex curve defined by $(z^2,z^3)=(0,0)$.

The mode expansion of the open strings connecting D7$_\pm$ to 
D7$_\pm$ itself is the same as that of the flat D7-branes. 
The fluctuation in $x^{0,1,2,3}$ and ${\bf T}^2$-directions 
have the same mode expansion also for the open strings connecting 
D7$_\mp$ to D7$_\pm$. However, the fluctuation in $z_2$ and $z_3$ 
directions have different mode expansion. 
The boundary condition is given by 
\begin{eqnarray}
 \partial_\sigma \Re \left( e ^{i\theta} (X^6 + i X^8) \right)  =  0,
  & \qquad &
 \partial_\sigma \Re \left( e ^{i\theta} (X^7 + i X^9) \right)  =  0,
  \\
 \partial_\tau \Im \left( e ^{i\theta} (X^6 + i X^8) \right)  =  0, 
  & \qquad &
 \partial_\tau \Im \left( e ^{i\theta} (X^7 + i X^9) \right)  =  0, 
\end{eqnarray}
at $\sigma = 0$ for the open strings starting from D7$_-$ and ending 
on D7$_+$. $e^{i\theta}$ is replaced by $e^{-i\theta}$ in the boundary 
conditions at $\sigma=\pi$.
The following mode expansion satisfies the above boundary conditions:
\begin{eqnarray}
 X^6+iX^8 & = & i \left(\frac{\alpha'}{2}\right)^{\frac{1}{2}} 
   \sum_{m \in \Z} 
    \left(  
       \frac{\alpha^{(6,8)}_{m+v}}{m+v} e^{-i (m + v)(\tau-\sigma)}
                                e^{-i\theta}
     + \frac{\beta^{(6,8)}_{m-v}}{m-v}  e^{-i (m - v)(\tau+\sigma)}
                                e^{-i\theta}
    \right), \\
 X^6-iX^8 & = & i \left(\frac{\alpha'}{2}\right)^{\frac{1}{2}} 
   \sum_{m \in \Z} 
    \left(  
       \frac{\beta^{(6,8)}_{m-v}}{m-v} e^{-i (m - v)(\tau-\sigma)}
                                e^{i\theta}
     + \frac{\alpha^{(6,8)}_{m+v}}{m+v}  e^{-i (m + v)(\tau+\sigma)}
                                e^{i\theta}
    \right), 
\end{eqnarray}
where $v \equiv 2\theta / \pi$, and 
oscillators $\alpha^{(6,8)}_{m+v}$ and $\beta^{(6,8)}_{m-v}$ satisfy 
$(\alpha^{(6,8)}_{m+v})^\dagger = \beta^{(6,8)}_{-m-v}$.
The mode expansion of $X^7\pm i X^9$ is exactly the same 
except that the oscillators $\alpha^{(6,8)}_{m+v}$ and 
$\beta^{(6,8)}_{m-v}$ 
are replaced by $\alpha^{(7,9)}_{m+v}$ and $\beta^{(7,9)}_{m-v}$.
The mode expansion of the world-sheet fermions is determined from 
that of the world-sheet bosons given above:
\begin{eqnarray}
 \psi^{(6,8)} =  \!\!\!\!\! \sum_{r \in \Z + \frac{1}{2}{\rm ~or~} \Z} 
      \psi^{(6,8)}_{r+v} e^{-i (r+v)(\tau-\sigma)} e^{-i\theta}, 
 & \qquad & 
 \tilde{\psi}^{(6,8)} = 
               \!\!\!\!\!\! \sum_{r \in \Z + \frac{1}{2}{\rm ~or~} \Z} 
    \bar{\psi}^{(6,8)}_{r-v}  e^{-i(r-v)(\tau+\sigma)} e^{-i\theta}, \\
 \bar{\psi}^{(6,8)} = 
               \!\!\!\!\!\! \sum_{r \in \Z + \frac{1}{2}{\rm ~or~} \Z} 
      \bar{\psi}^{(6,8)}_{r-v} e^{-i (r-v)(\tau-\sigma)}e^{i\theta}, 
 & \qquad & 
 \tilde{\bar{\psi}}^{(6,8)} = 
               \!\!\!\!\!\! \sum_{r \in \Z + \frac{1}{2}{\rm ~or~} \Z} 
    \psi^{(6,8)}_{r+v} e^{-i(r+v)(\tau+\sigma)} e^{i\theta}.
\end{eqnarray}
Here, $r \in \Z + 1/2$ for the NS sector and $\in \Z$ for the R sector.

The massless spectrum of the D7$_\pm$--D7$_\pm$ open string is quite 
simple --- that of the Yang--Mills theory on a 7+1-dimensional 
spacetime with 16 SUSY charges. 
The massless modes of the D7$_\mp$--D7$_\pm$ open string sector 
are localized at the intersection of D7$_-$ and D7$_+$, 
unless $\theta$ is an integral multiple of $\pi/2$.
A hypermultiplet in the bi-fundamental representation 
is obtained there; four scalar bosons of the hypermultiplet are 
$\psi^{(6,8)}_{-\frac{1}{2}+v}\ket{0;7_-7_+;NS}$, 
$\psi^{(7,9)}_{-\frac{1}{2}+v}\ket{0;7_-7_+;NS}$, 
$\bar{\psi}^{(6,8)}_{-\frac{1}{2}+v}\ket{0;7_+7_-;NS}$, and  
$\bar{\psi}^{(7,9)}_{-\frac{1}{2}+v}\ket{0;7_+7_-;NS}$, 
while fermions are obtained from the Clifford algebra of 
$\psi^{(2,3)}_0$, $\psi^{(4,5)}_0$, $\bar{\psi}^{(2,3)}_0$ and 
$\bar{\psi}^{(4,5)}_0$ in the R sector.
Here, we implicitly assume $0< \theta < \pi/4$ (i.e., $0< v < 1/2$) 
just to avoid technical details.

Now let us impose an orientifold projection associated with 
$\Omega R_{z^3} (-1)^{F_L}$. 
$R_{z^3}$ reflects the third complex plane, i.e., 
$R_{z^3}: z^3 \mapsto - z^3$. 
Thus, we have an O7-plane at $z_3 = 0$, and D7$_+$ is 
the orientifold mirror image of the D7$_-$ and vice versa.
Therefore, the Yang--Mills fields on D7$_-$ are identified with 
those on D7$_+$.
On the other hand, the D7$_\mp$--D7$_\pm$ open strings are mapped 
to themselves, not to each other.
Thus, the orientifold projection conditions are imposed:
\begin{eqnarray}
 \psi^{(6,8)/(7,9)}_{-\frac{1}{2}+v}\ket{0;i_-j_+;NS} & \sim &
 (\Omega R_{z^3}(-1)^{F_L}) 
   \psi^{(6,8)/(7,9)}_{-\frac{1}{2}+v}\ket{0;i_-j_+;NS}  \nonumber \\
& &  = - \psi^{(6,8)/(7,9)}_{-\frac{1}{2}+v}\ket{0;j_-i_+;NS}, 
\end{eqnarray}
for $0 < \theta < \pi/4$.
As a result, we have a hypermultiplet in the second-rank 
anti-symmetric-tensor representation.

An orbifold projection associated with a transformation 
\begin{equation}
  z^1 \mapsto e^{2i\alpha} z^1, \quad 
  z^2 \mapsto e^{-i\alpha} z^2, \quad 
  z^3 \mapsto e^{-i\alpha} z^3, 
\end{equation}
is now imposed, so that chiral matter content is obtained in 
the four-dimensional effective field theory.
$\alpha$ is an integral multiple of $2\pi/N$, 
when the orbifold group is ${\bf Z}_N$. 
The oscillators 
$\psi^{(6,8)}_{-\frac{1}{2}+v} \pm i \psi^{(7,9)}_{-\frac{1}{2}+v}$, 
and $\bar{\psi}^{(6,8)}_{-\frac{1}{2}+v} \pm 
      i\bar{\psi}^{(7,9)}_{-\frac{1}{2}+v}$ are multiplied by a phase  
$e^{\mp i \alpha}$ under the transformation. 
Suppose that the Chan--Paton matrix associated with the orbifold 
projection multiplies a phase $e^{i\beta}$ to D7$_-$--D7$_+$ states 
and a phase $e^{-i\beta}$ to D7$_+$--D7$_-$ states, then, 
two states 
$(\psi^{(6,8)}_{-\frac{1}{2}+v} + i \psi^{(7,9)}_{-\frac{1}{2}+v}) 
\ket{0;7_-7_+;NS}$ and 
$(\bar{\psi}^{(6,8)}_{-\frac{1}{2}+v} - 
      i\bar{\psi}^{(7,9)}_{-\frac{1}{2}+v}) \ket{0;7_+7_-;NS}$ 
satisfy the projection condition if $\alpha \equiv \beta$.
Likewise, two states in the R-sector D7$_-$--D7$_+$ open string 
are rotated by $e^{\pm i\alpha}$, because of the phase rotation 
of the oscillators $\psi^{(4,5)}_0$ and $\bar{\psi}^{(4,5)}_0$, 
and so are the two states in the R-sector D7$_+$--D7$_-$ open string.
Thus, one state from D7$_-$--D7$_+$ string and one state from 
D7$_+$--D7$_-$ survive the orbifold projection condition 
when $\alpha \equiv \beta$.
These two states from the NS sector and two states from the R sector 
form a chiral multiplet of the ${\cal N}$ = 1 SUSY of 
four-dimensional spacetime.
This multiplet is a half of the hypermultiplet in the anti-symmetric 
representation, and the other half (i.e., the chiral multiplet 
in the conjugate representation) is projected out.

\section{\label{sec:FamilyK}Global and Local Formulae \\
of the Number of Chiral Families}

Quarks and leptons arise at the intersection of the two stacks 
of the D7-branes.
The number of families in D = 4 effective theory is given 
by formula (\ref{eq:NB},\ref{eq:NF}), which only uses 
quantities defined locally around the intersection.
Incidentally, the number of chiral multiplets 
in D = 4 effective theory is obtained also through a formula 
in \cite{Dquint}, which is given in terms of vector bundles on the 
whole Calabi--Yau 3-fold.
The purpose of this appendix is to show the equivalence between 
them explicitly.

The Ramond--Ramond charges of D-branes are classified as elements 
of even-dimensional cohomology groups of the Calabi--Yau 3-fold $X$ 
\cite{Mukai,Iflow,HvMr,Iflow-curved,MM,MSSSS}:
\begin{equation}
 v_X(\Sigma,E_\Sigma)  =  
  {\rm ch}((i_\Sigma)_! E_\Sigma \otimes {\cal L}_B^{-1}) 
  \sqrt{\hat{A}(TX)}  \in H^{\rm even}(X),
\label{eq:RRchargeA}
\end{equation}
for a D-brane wrapped on a cycle 
$i_\Sigma : \Sigma \hookrightarrow X$ with Chan--Paton bundle 
$E_\Sigma$ on it, or equivalently, 
\begin{equation}
 v_X(\Sigma,E_\Sigma)  =  
  (i_\Sigma)_* \left(
     {\rm ch}(E_{\Sigma}\otimes {\cal L}_B^{-1}) 
     \sqrt{\frac{\hat{A}(T\Sigma)}
                {\hat{A}({\cal N}_{\Sigma | X})}}
                     \right) \in H^{\rm even}(X).
\label{eq:RRchargeB}
\end{equation}
Here, $(i_\Sigma)_*$ and $(i_\Sigma)_!$ are the Thom isomorphism 
for cohomology and its analogue for vector bundles, respectively, 
associated with the embedding $i_\Sigma$.
The pairing of the Ramond--Ramond charges is defined by 
\begin{equation}
 I_X((\Sigma_1,E_{\Sigma_1}),(\Sigma_2,E_{\Sigma_2})) 
= \int_X v_X(\Sigma_1,E_{\Sigma_1}) \raisebox{2mm}{$\vee$} \wedge 
         v_X(\Sigma_2,E_{\Sigma_2}),
\end{equation}
where $v_X(\Sigma_1,E_{\Sigma_1})\raisebox{2mm}{$\vee$}$ 
is obtained from 
$v_X(\Sigma_1,E_{\Sigma_1})$ by multiplying $(-1)^q$ to 
$2 q$-dimensional cohomology, or equivalently,  
\begin{eqnarray}
 I_X((\Sigma_1,E_{\Sigma_1}),(\Sigma_2,E_{\Sigma_2}))
  & = & \int_X {\rm ch} \left[ 
       \left( (i_{\Sigma_1})_! E_{\Sigma_1} \otimes {\cal L}_B^{-1}
       \right)^* \otimes 
       \left( (i_{\Sigma_2})_! E_{\Sigma_2} \otimes {\cal L}_B^{-1} 
       \right) 
                    \right] \hat{A}(TX) \nonumber \\
 & = & {\rm index}_X \Dsl_{
       \left( (i_{\Sigma_1})_! E_{\Sigma_1} \otimes {\cal L}_B^{-1}
       \right)^* \otimes 
       \left( (i_{\Sigma_2})_! E_{\Sigma_2} \otimes {\cal L}_B^{-1} 
       \right) }. 
\label{eq:pairing}
\end{eqnarray}
Thus, it is given by the number of fermion zero modes 
on the Calabi--Yau 3-fold $X$. 
The number of fermion zero modes on $\Sigma_1 \cdot \Sigma_2$ 
obtained in (\ref{eq:NF}) 
is equal to the number of zero modes obtained above :
\begin{eqnarray}
& & I_X((\Sigma_1,E_{\Sigma_1}),(\Sigma_2,E_{\Sigma_2})) =  
   - I_X((\Sigma_2,E_{\Sigma_2}),(\Sigma_1,E_{\Sigma_1})) 
      \nonumber \\
& = & \!\!\!\!\!
       \int_X (i_{\Sigma_2})_* \left(
   {\rm ch}(E^*_{\Sigma_2}\otimes {\cal L}_B) 
   \sqrt{\frac{\hat{A}(T\Sigma_2)}
              {\hat{A}({\cal N}_{\Sigma_2 | X})}}\right) \wedge 
            (i_{\Sigma_1})_*  \left(
   {\rm ch}(E_{\Sigma_1}\otimes {\cal L}_B^{-1}) 
   \sqrt{\frac{\hat{A}(T\Sigma_1)}
              {\hat{A}({\cal N}_{\Sigma_1 | X})}}\right) 
                   \nonumber \\
& = & \int_{\Sigma_1 \cdot \Sigma_2}
     (({i_{\Sigma_1}})|_{\Sigma_1 \cdot \Sigma_2})^*
        \left(
   {\rm ch}(E^*_{\Sigma_2}\otimes {\cal L}_B) 
   \sqrt{\frac{\hat{A}(T\Sigma_2)}
              {\hat{A}({\cal N}_{\Sigma_2 | X})}}
        \right) \wedge  \nonumber \\
& & \qquad \qquad \qquad 
     (({i_{\Sigma_2}})|_{\Sigma_1 \cdot \Sigma_2})^* 
        \left(
   {\rm ch}(E_{\Sigma_1}\otimes {\cal L}_B^{-1}) 
   \sqrt{\frac{\hat{A}(T\Sigma_1)}
              {\hat{A}({\cal N}_{\Sigma_1 | X})}}\right)  \nonumber \\
& = & \!\!\!\! \int_{\Sigma_1 \cdot \Sigma_2} \!\!\!\!\!\!
   {\rm ch}\left( 
     (({i_{\Sigma_1}})|_{\Sigma_1 \cdot \Sigma_2})^* 
          E^*_{\Sigma_2} \otimes 
     (({i_{\Sigma_2}})|_{\Sigma_1 \cdot \Sigma_2})^* 
          E_{\Sigma_1} 
            \right) \hat{A}(T(\Sigma_1 \cdot \Sigma_2)) \nonumber \\
 & = & N_F.
\end{eqnarray}

The Ramond--Ramond charge of an orientifold plane\footnote{Newly added 
in version 2, March, 2006.} $W \subset X$ 
is given by (e.g. \cite{Oplane,Toronto})
\begin{equation}
 v_X(W) = - (i_W)_* \left[ 2^{\dim_{\bf R}W - \dim_{\bf C}X}
   \sqrt{\frac{L\left(\frac{1}{4}TW \right)}
              {L\left(\frac{1}{4}N_{W|X}\right)}
        } \right].
\end{equation}
The multiplicities of chiral multiplets in the anti-symmetric and 
symmetric representations are given, respectively, by 
\begin{eqnarray}
 {\rm a.sym}: & & \frac{1}{2}\left( 
   I_X (v_X(\Sigma_1,E_{\Sigma_1}), v_X(\Sigma_{1'},E_{\Sigma_{1'}}) )
  - I_X (v_X(\Sigma_1,E_{\Sigma_1}) , v_X(W) ) \right), \\
 {\rm sym}: & & \frac{1}{2}\left( 
   I_X (v_X(\Sigma_1,E_{\Sigma_1}), v_X(\Sigma_{1'},E_{\Sigma_{1'}})  )
  + I_X (v_X(\Sigma_1,E_{\Sigma_1}) , v_X(W) ) \right).
\end{eqnarray}
In an application to a system of intersecting D7-branes and 
an O7-plane, $2^{\dim_{\bf R}W - \dim_{\bf C}X}=2$. 
They are also expressed by integration over the intersection curve 
$\Sigma_1 \cdot \Sigma_{1'} = \Sigma_1 \cdot W$. Note that 
$\hat{A}$ and $L$ do not contribute in integration over complex 
curves, since they do not have 2-form components.
Thus, the multiplicities are 
\begin{eqnarray}
 {\rm a.sym.}: & & \frac{1}{2}\left( 
  \int_{\Sigma_1 \cdot \Sigma_{1'}} 
  \left(\frac{F_{\Sigma_1}}{2\pi}- \frac{F_{\Sigma_{1'}}}{2\pi}\right)
+ \int_{\Sigma_1 \cdot W} 2
\left(\frac{F_{\Sigma_1}}{2\pi}-\frac{B}{(2\pi)^2\alpha'}\right)
\right), \\
 {\rm sym.}: & & \frac{1}{2}\left( 
  \int_{\Sigma_1 \cdot \Sigma_{1'}} 
  \left(\frac{F_{\Sigma_1}}{2\pi}- \frac{F_{\Sigma_{1'}}}{2\pi}\right)
- \int_{\Sigma_1 \cdot W} 2
  \left(\frac{F_{\Sigma_1}}{2\pi}-\frac{B}{(2\pi)^2\alpha'}\right)
\right).
\end{eqnarray}
SInce the orientifold projection condition on the intersection 
curve $\Sigma_1 \cdot \Sigma_{1'} = \Sigma_1 \cdot W$ is 
\begin{equation}
 \left(\frac{F_{\Sigma_1}}{2\pi}-\frac{B}{(2\pi)^2\alpha'}\right)= - 
 \left(\frac{F_{\Sigma_{1'}}}{2\pi}-\frac{B}{(2\pi)^2\alpha'}\right),
\end{equation}
no chiral multiplets in the symmetric representation arise on the 
normal orientifold planes, and the multiplicity of the anti-symmetric 
representations is given by (\ref{eq:Nasym}).

\end{document}